# Quantum flicker noise in atomic and molecular junctions

Ofir Shein-Lumbroso[1], Junjie Liu[2,*], Abhay Shastry[2,*], Dvira Segal[2,3] and Oren Tal[1]

[1]Department of Chemical and Biological Physics, Weizmann Institute of Science, Rehovot 7610001, Israel
[2]Department of Chemistry and Centre for Quantum Information and Quantum Control, University of Toronto, 80 Saint George St., Toronto, Ontario, M5S 3H6, Canada
[3]Department of Physics, 60 Saint George St., University of Toronto, Toronto, Ontario, Canada

[*]Equal contributions

**Abstract:**
We report on a quantum form of electronic flicker noise in nanoscale conductors that contains valuable information on quantum transport. This noise is experimentally identified in atomic and molecular junctions, and theoretically analyzed by considering quantum interference due to fluctuating scatterers. Using conductance, shot noise, and flicker noise measurements, we show that the revealed quantum flicker noise uniquely depends on the distribution of transmission channels, a key characteristic of quantum conductors. This dependence opens the door for the application of flicker noise as a diagnostic probe for fundamental properties of quantum conductors and many-body quantum effects, a role that up to now has been performed by the experimentally less-accessible shot noise.

**Main:**

Flicker noise, sometimes called $1/f$ noise, is typically regarded as the most ubiquitous form of noise in nature (e.g., [1-4]). It is also experimentally accessible and widely studied. However, shot noise is in fact the dominant form of noise used for fundamental characterization of quantum transport and related many-body effects. This is despite the challenges involved in measuring shot noise due to its small signal with respect to other noise contributions. Specifically, the combination of electronic conductance and electronic shot noise measurements in quantum coherent conductors has been used extensively to extract information on quantum transport. For example, such measurements play a central role in the analysis of the fractional quantum Hall effect [5,6], Kondo interaction [7,8], spin-polarized quantum transport [9-14], electron-vibration interaction [15-18], as well as in revealing the influence of local atomic structure on the conductance of atomic and molecular junctions [19-24]. Electronic shot noise is a useful source for information because it depends on the distribution of transmission channels, which determines quantum transport in the framework of Landauer formalism [25]. These channels are the transmission modes available for wavelike electrons crossing a quantum coherent conductor in some analogy to electromagnetic wave modes in a waveguide. For $eV >> k_B T$ the dependence of shot noise on transmission channels is given by [12,25] $S_{SN} = 2eIF$, where $F = [\sum_i \tau_i (1 - \tau_i)]/\sum_i \tau_i$ is the Fano factor. Here, $e$ is the electron's charge, $V$ is the applied voltage across a conductor, $k_B$ is the Boltzmann's factor, $T$ is the temperature, $I$ is the current, and $\tau_i$ is the transmission probability at the Fermi energy of the $i^{th}$ transmission channel. Together with the distinct dependence of the conductance $G$ on the transmission channels [25], $G = G_0 \sum_i \tau_i$, ($G_0 = 2e^2/h$ is the conductance quantum and $h$ is the plank's constant) the two quantities of shot noise and conductance can provide information on the distribution of transmission channels in quantum conductors and as a result, allows the explorations of many-body interactions in quantum devices.

Electronic flicker noise has been measured in a variety of nanoscale systems (e.g., [26-34]), including atomic and molecular junctions [35-44]. However, the quantum nature of flicker noise as manifested in the relation between this noise and the distribution of transmission channels has not been examined experimentally or theoretically, despite the important role of these channels. Here, we reveal a quantum version of flicker noise with a unique dependence on the channels' transmission probabilities, distinct from the behavior of conductance and shot noise. We use the break junction technique [45] (Fig. 1a) to jointly measure conductance, flicker noise, and shot noise in an ensemble of atomic and molecular junctions based on gold (Au) and hydrogen. The relation between the measured flicker noise and transmission channels is analyzed with the aid of a model based on quantum interference in the presence of fluctuating scatterers located in the vicinity of the junction (illustrated in Fig. 1a, inset). Based on the measured flicker noise, shot noise, and conductance, we perform a transmission channel analysis [46] reaching a higher accuracy than when merely using the latter two, as is commonly done. Typically, flicker noise is more experimentally accessible than shot noise. Therefore, the combination of flicker noise and conductance measurements, or, alternatively flicker noise, conductance, and shot noise, can promote a more widespread (in the former case), and more accurate (in the latter case) analysis of transmission channels in quantum conductors. More broadly, beyond the framework of Landauer transport, we anticipate that the revealed nature of flicker noise in quantum conductors would provide useful information on Kondo systems, superconducting point contacts, fractional quantum Hall devices, and electron-phonon interaction in atomic scale junctions, in analogy to the diagnostic role of shot noise in such cases.

We treat transport of electrons in an atomic-scale junction at low temperature using a quantum coherent wave picture. The junction is separated into three regions, as depicted in Fig. 1a, inset: (I) The central region (denoted as C) of the junction is of atomic dimensions and it supports ballistic transport. (II) To the left and right of the atomic constriction region we identify the "interface zones", which extend within the coherent mean free path.

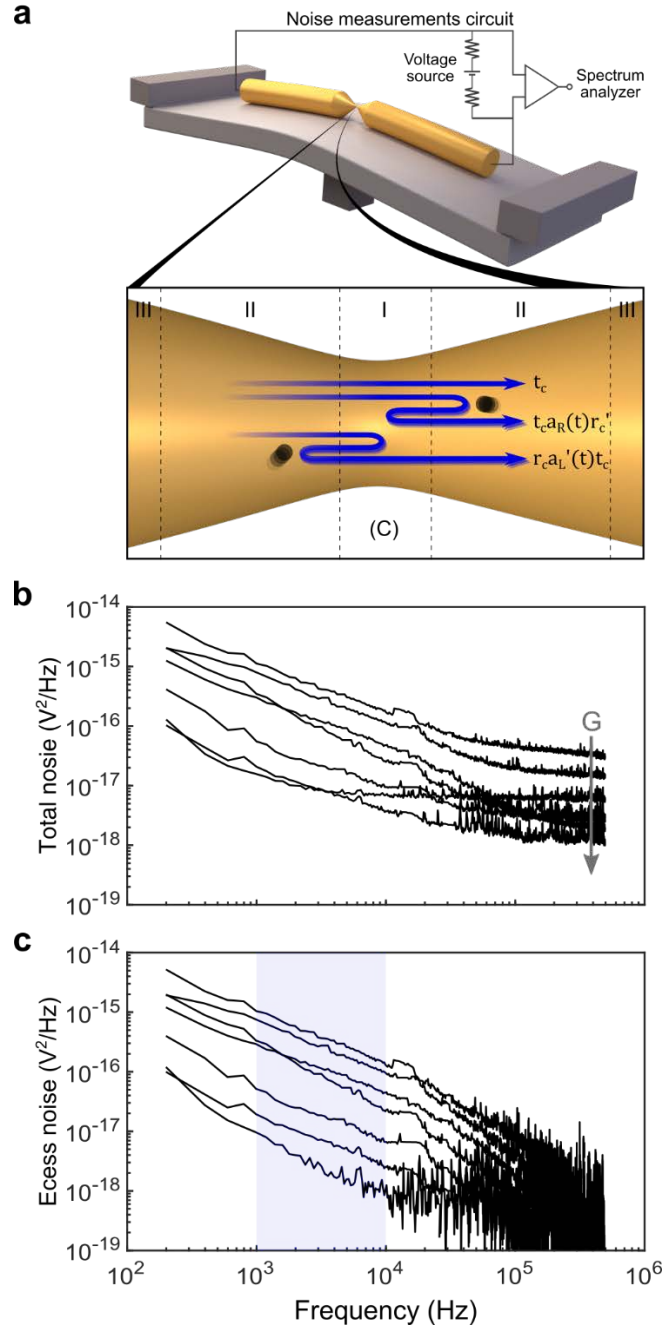

FIG. 1. (a) Illustration of a break junction setup and the measurement circuit. Inset: illustration of the flicker noise model. Conducting electrons experience coherent scattering in the atomic-scale junction (region I denoted as C) as well as by elastic scatterers located near the contact (in regions II), resulting in a quantum interference term that contributes to the junction's overall transmission. Fluctuations of the scatterers cross-section in region II generate a flicker noise with quantum characteristics. Regions III represent the rest of the semi-infinite electrodes. (b) Measured total noise as a function of frequency for seven Au/hydrogen junctions with conductance between 0.55 $G_0$ and 7.32 $G_0$ top to bottom. (c) Excess noise (practically flicker noise) as a function of frequency after the subtraction of thermal noise and shot noise. The circuit spurious output voltage noise and amplifier input current noise were subtracted, and the RC low-pass suppression was corrected using the transfer function of thermal noise (see Supplemental Material).

Scattering processes within these regions are assumed to be elastic, and are treated using the coherent scattering approach. Fluctuating defects in the interface zones result in changes to the cross section for scattering of electrons - yet on a timescale long relative to the electron transmission time through the junction. These dynamical defects are responsible for the physics of the flicker noise in our system. (III) Away from the two interface zones, beyond the phase-coherent and elastic mean free path, the rest of the structure is treated as an ideal metal. We assume that the electron occupation in the bulk is described by the Fermi-Dirac distribution function at a given temperature and Fermi level.

Since the dynamics of defects occur on a timescale much longer than that of electron transport through the contact, we derive an expression for the flicker noise that depends on the distribution of transmission channels in the framework of Landauer-Büttiker formalism [25]. To obtain the transmission function in each channel, we follow Ludoph $et$ $al.$ [47,48], taking into account the interference of incoming electrons with a wave component that is reflected due to scattering with defects at the interface zones. However, in our model scatterers have a dynamic scattering cross-section. The derived expression:

$$S_f(\omega) = S \sum_i \tau_i^2 (1 - \tau_i), \qquad (1)$$

provides a quantum version of flicker noise, using $S \equiv 2G_0^2 V^2 \Phi(\omega)$. As can be seen in the Supplemental Materiel, Eq. (1) was obtained after ($i$) neglecting correlations between different channels and different electrodes, and ($ii$) assuming a power spectrum of reflection amplitudes due to defects in region II, $\Phi(\omega)$, which does not depend on the channel index. In other words, we assume that in the multi-channel case electrons in different channels are affected by the same defect configurations. The power spectrum $\Phi(\omega)$ and the transmission probability $\tau_i$ are evaluated at the Fermi level. Eq. (1) connects between the measured flicker noise and the microscopic picture of transmission channels, which is the focus of our analysis. Note that this equation can be generalized by allowing the transmission and reflection processes of the central region to be time-dependent as well (Supplemental Material).

Figure 1b shows the total noise measured for several Au atomic-scale junctions in the presence of hydrogen (Au/hydrogen junctions). The introduction of hydrogen gas into the cold junction ($4.2 \ K$) allows us to study flicker noise in junctions with a wide range of conductance values, also below the $\sim 1 \ G_0$ conductance of a single Au atomic contact [47]. To extract the flicker noise contribution, we follow the technical procedure described in the Supplemental Material and Refs. [49,50]. Briefly, we subtract unwanted contributions of circuit output voltage noise and amplifier current input noise. Furthermore, we correct the signal suppression by the setup RC filtering (R-resistance, C-capacitance). Finally, we probe the thermal and shot noises at a high frequency range, where the flicker noise is negligible (280-290 kHz), and subtract their contributions from the total measured noise to reveal the flicker noise component. Examples for the obtained noise using this process are presented in Fig. 1c. The measured flicker noise depends on the frequency with a power that scatters around 1 ($i.e.,$ $\alpha \cong 1$, using Hooge's expression [47]: $S_f \sim 1/f^\alpha$), and shows a typical quadratic dependence on applied voltage in the examined $mV$ range (see Supplemental Material). Finally, we integrate over the noise in the range of $10^3$-$10^4$ Hz (shaded in Fig. 1c).

The integrated noise $S_f$ for 623 junctions with different conductance values is presented Fig. 2a. In what follows, we analyze the integrated flicker noise data in view of Eq. (1). As a first step, we focus on the $S$ prefactor. $\Phi(\omega)$ may vary between junction to junction since is it sensitive to the details of the fluctuating scatterers that can be different for distinct junctions. As a result, $S$ has a range that can be characterized by $S_{min}$ and $S_{max}$. To find these values, we focus on noise data between 0.1 and 1 $G_0$, for which former shot noise measurements on Au\hydrogen junctions [50] revealed that the conductance was dominated by a single transmission channel, while a minor contribution from a second channel (or more in rare cases) was found when the conductance approached 1 $G_0$, probably due to direct Au-Au tunneling.

Fig. 2b illustrates that below 2/3 $G_0$, the expression $\sum_i \tau_i^2(1 - \tau_i)$ is maximal if a single channel contributes (red curve). In contrast, between 2/3 and 1 $G_0$ a single channel leads to minimal flicker noise (red curve), compared to multi-channel junctions. As a result, we can fit in Fig. 2a $S\tau^2(1 - \tau)$, with $\tau = G/G_0$ to the lowest data points between 2/3 and 1 $G_0$ to find $S_{min}$ and repeat the fitting for the highest data points below 2/3 $G_0$ to obtain $S_{max}$ (red curves). The semitransparent red region in Fig. 2a describes the expected flicker noise for a single transmission channel with $S$ between $S_{max}$ and $S_{min}$ due to variations in the scatterers' distribution for different junctions. The data spread above the semitransparent red region increases as the conductance approaches 1 $G_0$. This characteristic evolution is ascribed to the contribution of more than a single transmission channel (Supplemental Fig. S3). The upper limit of flicker noise according to Eq. (1) is depicted by the upper boundary of the gray area in Fig. 2a. Along this curve, which to a good approximation is proportional to $G/4$, the conductance is made of equal transmission probabilities. Namely, for $N$ channels: $\tau_1 = \tau_2 = \cdots = \tau_N = G/(NG_0)$ (Supplemental Material).

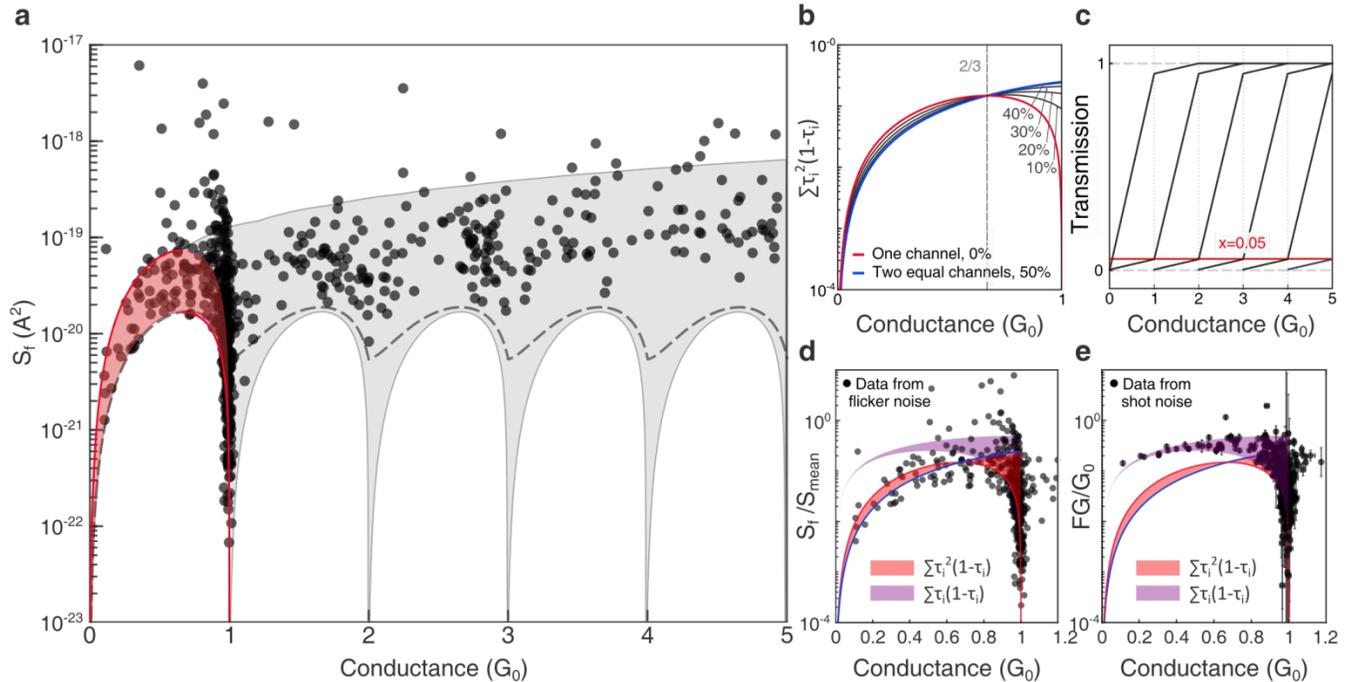

FIG. 2. (a) Flicker noise integrated in the range $10^3$-$10^4$ $Hz$ as a function of conductance. Each data point (black) is measured for a different Au/hydrogen junction realization. The upper red curve is a fit of $S_f = S_{max}\tau_1^2(1 - \tau_1)$ between 0 to 2/3 $G_0$ and the lower red curve is a fit to $S_f = S_{min}\tau_1^2(1 - \tau_1)$ between 2/3 $G_0$ and 1 $G_0$ (only data below $10^{-19}$ $A^2$ where considered). The gray area presents the allowed flicker noise values based on Eq. (1). However, its lower boundary is relevant only for an ideal sequential opening of channels. The dashed black curve provides the lower boundary for flicker noise for a non-ideal sequential opening of channels as presented in c. (b) $\sum_i \tau_i^2(1 - \tau_i)$ as a function of conductance for one and two transmission channels. This expression is maximal between 0 to 2/3 $G_0$ and minimal between 2/3 to 1 $G_0$ for a single channel. (Supplemental Material). (c) Model for channel transmission probabilities as a function of conductance for a non-ideal sequential opening of channels, based on Ref. [19]. Here, more than one channel is partially open for a given conductance, where the transmissions increase linearly and the consequential mode contributes 5% (x=0.05) at integer conductance. (d) $S_f / S_{mean}$ as a function of conductance. Semitransparent red and purple areas are the ensembles of $\sum_i \tau_i^2(1 - \tau_i)$ and $\sum_i \tau_i(1 - \tau_i)$, respectively for all possible values of $\tau_1$ and $\tau_2$ for conductance below $1G_0$. (e) Similar to (d), but for $FG/G_0$. The blue and red curves in (d)-(e) correspond to $\sum_i \tau_i^2(1 - \tau_i)$ with two equally opened channels and a single-channel, respectively. The data presented in (a) were converted from voltage power units ($V^2$) to current power units ($A^2$) by dividing each value with the square of the corresponding junction's resistance. The data error range is comparable or smaller than the diameter of the symbols.

The appearance of some data points above this curve (e.g., 7% of the data points between 0.9 and 1.0 $G_0$ are located above the gray region) can indicate on additional noise contributions that are not described by Eq. (1), such as conductance fluctuations that stem from junction instability. The bottom boundary of the gray region presents the lower limit of flicker noise based on Eq. (1) for the case of sequential opening of channels, as expected for an ideal quantum point contact [52]. In this case, as the conductance increases, it is given by one channel up to 1 $G_0$ with transmission probability of $\tau_1 = G/G_0$. Then, a second channel is opened, while the first channel remains fully open with $\tau_1 = 1$ and $\tau_2 = G/G_0 - 1$ up to 2 $G_0$, etc. However, for Au atomic contacts (with or without hydrogen) the opening of channels when the conductance increases is not fully sequential. Namely, before a given channel is fully open ($\tau_i = 1$), another channel or more are already partially opened (e.g., $0 < \tau_{i+1} < 1$) [19,46,53]. The essence of this situation was captured by H. E. van den Brom *et al.* [19], assuming a simple model for non-ideal sequential opening of channels, as illustrated in Fig. 2c. When considering the presented channel evolution in Fig. 2c, Eq. (1) yields a lower limit for flicker noise as a function of conductance seen as a dashed black curve in Fig. 2a. This curve describes better the minimal values of the measured flicker noise. Note that in reality, the number of partially open channels for a given conductance slightly increases as the conductance increases [46,53,54]. Consequently, the lower boundary for the measured flicker noise should slightly increase at higher conductance, and therefore deviate from the dashed curve, as indeed seen in Fig. 2a above ~3 $G_0$.

By combining measurements of flicker noise and shot noise on the same junctions, we check if the two types of probed noise indeed reveal distinctive dependence on the distribution of transmission channels, as expected by the theoretical treatment. This dependence can be conveniently expressed as $S_f/S = \sum_i \tau_i^2(1 - \tau_i)$ for the flicker noise and $FG/G_0 = \sum_i \tau_i(1 - \tau_i)$ for shot noise. Fig. 2d and e present the measured $S_f/S$, and $FG/G_0$, respectively (semitransparent black dots). Note that the Fano factor is determined by the dependence of the measured shot noise on applied current [7,33,42]. For the prefactor $S$ of the flicker noise, we use $S_{mean} = (S_{min} + S_{max})/2$. On top of these normalized experimental data, we present the calculated $\sum_i \tau_i^2(1 - \tau_i)$ in red and $\sum_i \tau_i(1 - \tau_i)$ in purple, assuming up to two transmission channels with all possible combinations of $\tau_1$ and $\tau_2$ that satisfy $G = G_0(\tau_1 + \tau_2)$. As can be seen, the center of the measured $S_f/S$ in Fig. 2d is well described by $\sum_i \tau_i^2(1 - \tau_i)$, as expected when using $S_{mean}$, and the measured $FG/G_0$ (Fig. 2d) is captured by $\sum_i \tau_i(1 - \tau_i)$ (semitransparent purple). This analysis verifies that the probed flicker noise and shot noise are two independent functions of the channel's transmission probabilities.

The distinct dependence of conductance and shot noise on transmission channels has been employed to reveal information on the number of transmission channels and their respective transmission probabilities. This approach allowed detecting spin-dependent quantum transport, relating transmission channels to the local electronic structure of atomic and molecular junctions, and understanding the effect of electron-vibration interaction on quantum conductance [9-24]. However, using two independent equations (for conductance and shot noise) discloses analytically only two or less transmission probabilities. This limitation is partially lifted by adopting numerical approaches that provide information on the transmission probabilities of more than two channels, yet with the cost of reduced accuracy [46].

Thanks to the characteristic dependence of flicker noise on the channel distribution, which is different than the ones of conductance and shot noise, we can now utilize flicker noise for channel analysis. Fig. 3a presents the Fano factor extracted from shot noise measurements of Au/hydrogen junctions as a function of conductance, where each $(F, G)$ data point was measured on a different junction. To examine the use of flicker noise in numerical channel analysis, which is usually based merely on conductance and shot noise, we focus on two data points labeled as I and II in Fig. 3a. Fig. 3b and c, present the transmission probabilities of the four most dominant transmission channels. The blue distribution for each channel indicates the possible range of transmission probabilities $\tau_i$, based only on shot noise and conductance analysis (the relevant values are given in Table I). For example, the second channel of junction I, has a transmission probability between $0.06 \leq \tau_2 \leq 0.26$, where the uncertainty comes from the application of two equations to obtain information on four transmission probabilities [46]. The red distributions are calculated based on flicker noise, shot noise and conductance data (Table I). Proceeding with our example, this analysis yields a transmission probability for the second channel of junction I in the range of $0.19 \leq \tau_2 \leq 0.26$, namely with a reduced uncertainty. Similarly, the analysis of other transmission channels in Fig. 3b and c show that the adoption of flicker noise data on top of conductance and shot noise data leads to improved accuracy in the analysis (generally, the accuracy can be better or equal).

Shot noise measurements are typically very demanding due to relatively low signal. In contrast, flicker noise (in combination with conductance) measurements offer a more experimentally-accessible approach for transmission channels analysis in quantum conductors. In fact, in different schemes of shot noise measurements, the contribution of flicker noise is measured unintentionally, typically as an unwanted component that should be identified and removed. In these cases, flicker noise can serve fruitfully, together with shot noise, as a mean for a more accurate analysis of transmission channels with no special setup adjustments. Note that in the experimental approach adopted here, we deform the electrodes' apices between consecutive junction realizations by crashing the electrodes against each other. This is done to probe the span of possible junction geometries and the spread of the ensuing flicker noise. As a result, a wide range of the prefactor $S$ is expected due to variations in the properties and distribution of fluctuating scatterers located in the electrode apices. A moderate or no squeezing of the electrodes against each other between consecutive junction realizations can minimize variations in $S$ by preserving the characteristics of fluctuating scatterers near the junction. This mode of work can be adopted for achieving an accurate channel analysis, on account of sampling a large variety of junction configurations. Clearly, this issue is less relevant for quantum conductors beyond break junction systems. We note that ensemble-averaged conductance fluctuations measured for thousands of atomic scale junctions have shown a similar *collective* dependence on transmission channels as found here for flicker noise due to a similar origin [46,47]. However, while this experimental approach probes ensemble-averaged properties of channel distributions, it cannot be utilized for channel analysis in individual quantum conductors, in contrast to conductance, shot noise and flicker noise measurements.

At the tunneling limit for a single channel, flicker noise can be used to extract the charge of quasiparticles, thus providing an independent probe that complements similar shot noise based analysis of quasiparticles and electron-electron interactions. Electron-phonon (or electron-vibration) interaction has been studied in atomic and molecular junctions for more than two decades [55-58]. Specifically, an extensive theoretical work addressed the influence of phonon-activation on conductance [59-68] and shot noise [69-74] for off-resonance transport and weak electron-phonon coupling. In these conditions and in view of Eq. (1), flicker noise should also be affected by electron-phonon interaction. For a transmission probability smaller (larger) than 2/3, flicker noise is expected to increase (decrease) when a phonon mode is activated (see Supplemental Material). This response can provide a complementary independent analysis tool, beyond conductance and shot noise for the study of electron-phonon interaction in atomic contacts, atomic chains, and molecular junctions.

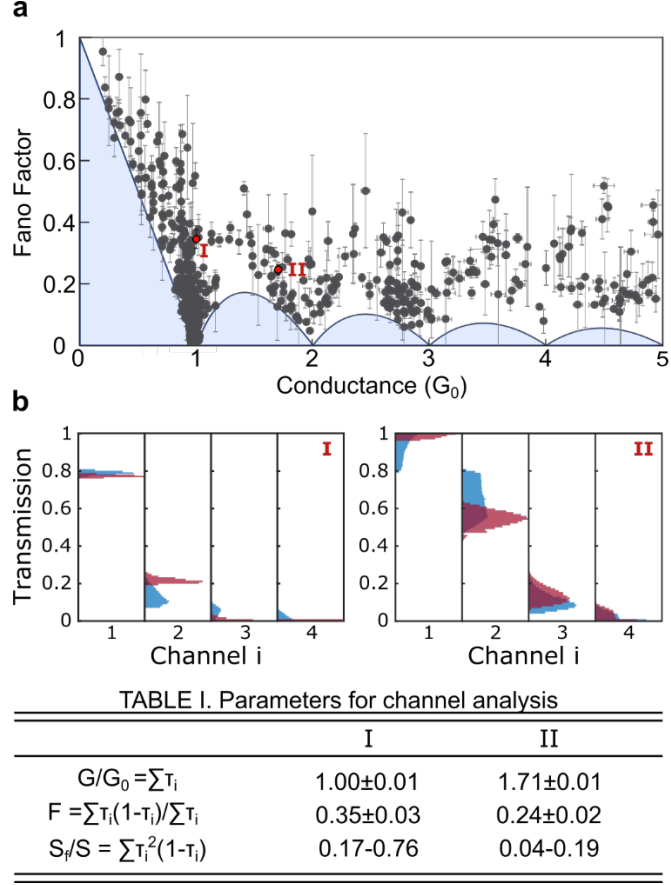

# Supplemental Material: Quantum flicker noise in atomic and molecular junctions

Ofir Shein-Lumbroso,[1] Junjie Liu*,[2] Abhay Shastry*,[2] Dvira Segal,[2,3] and Oren Tal[1]

[1] *Department of Chemical and Biological Physics, Weizmann Institute of Science, Rehovot 7610001, Israel*
[2] *Chemical Physics Theory Group, Department of Chemistry and Centre for Quantum Information and Quantum Control,
University of Toronto, 80 Saint George St., Toronto, Ontario, M5S 3H6, Canada*
[3] *Department of Physics, 60 Saint George St., University of Toronto, Toronto, Ontario, Canada M5S 1A7*
(Dated: July 14, 2021)

*equal contribution

In this Supplemental material, we provide additional details on the experimental procedure (Part I on sample fabrication, measurements, data analysis), and derive the expression for flicker noise in atomic-scale junctions (Part II on theoretical derivations). In the latter, we focus on the relationship between the flicker noise and the transmission probability through the atomic-scale junction, and derive bounds on the flicker noise in terms of the transmission.

## CONTENTS

# PART I: EXPERIMENT

## S1. ADDITIONAL INFORMATION ON MEASUREMENTS

### A. Sample fabrication in a break-junction setup

We used a mechanically-controllable break-junction setup located within a cryogenic chamber to produce atomic and molecular junctions and characterize them, as described in Ref.[1]. Briefly, the chamber is pumped to $10^{-5}$ mbar and cooled by liquid helium to about 4.2 K. This setup is placed in a Faraday cage to facilitate flicker noise and shot noise measurements. The sample is made of a notched Au wire (99.99%, 0.1 mm diameter, 25 mm length, Goodfellow), which is anchored on a flexible substrate (0.76 mm thick insulating Cirlex film). A three-point bending mechanism is then used to bend the substrate and break the wire at the notch (Fig1a in the main text). The wire is broken in cryogenic vacuum to expose two ultra-clean atomically sharp tips that serve as the junction's electrodes. The breaking process is controlled by a piezoelectric element (PI P-882 PICMA), connected to a Piezomechanik SVR 150/1 piezo driver, which is driven by a 24-bit NI-PCI4461 data acquisition (DAQ) card. These components provide fast control over the distance between the two tips with sub-angström resolution. To form molecular junctions, pure hydrogen gas (99.999%, Gas Technologies) was admitted from an external molecular source to the cold junction via a stainless steel capillary. The formation of Au/hydrogen junctions was monitored during the admission process by recording deviations from the typical conductance of bare Au. Following the formation of molecular junctions, the admission of hydrogen was stopped.

### B. Conductance measurements

To measure conductance, direct-current (d.c.) was measured on atomic and molecular junctions with a fixed inter-electrode distance as a function of applied voltage. The voltage was provided by a NI-PCI4461 DAQ card, and the resulting current was amplified by a current preamplifier (SR570) and recorded by the DAQ card. Following each junction analysis, the exposed atomic tips were pushed back into contact until the conductance reaches a value of at least 50 $G_0$, in order to ensure that the data consist of a statistical variety of atomic scale junctions with different geometries. The instruments were connected to a quiet ground, and were optically isolated from a control computer outside the Faraday cage. The amplifiers were powered by batteries to avoid noise injection from power lines. Additionally, an $RC$ filter (where $R$ is resistance and $C$ is capacitance) was connected between the piezo driver and the piezoelectric element to minimize possible excitation of mechanical noise coupled to the junction through the piezoelectric element.

### C. Flicker noise analysis

Noise measurements were performed on atomic and molecular junctions using the circuit described in Fig. 1a of the main text. The sample was current-biased by a Yokogawa GS200 SC voltage source connected to the sample through two 0.5 MΩ or 1 MΩ resistors located near the sample. The resulting voltage noise was amplified by a custom-made differential low-noise amplifier and analyzed via a NI PXI-5922 DAQ card using a LabView implemented fast Fourier transform (FFT) analysis. The custom-made amplifier was calibrated by the thermal (Johnson–Nyquist) noise that is generated in a set of well-characterized resistors embedded in liquid nitrogen. A power spectrum between 0.25 kHz and 300 kHz was measured via the DAQ card using the mentioned LabView FFT analysis and averaged 1,000 times. Fig. S1a presents the measured total voltage noise as a function of frequency.

To extract the flicker noise contribution, we follow the technical procedure described next. The unwanted contribution of the setup output voltage noise is measured for a shorted (mechanically squeezed) junction and subtracted. Then, the signal suppression by the setup's circuit RC filtering is characterized using thermal noise analysis and corrected. Next, the unwanted amplifier input current noise contribution is identified and subtracted, yielding the data presented in Fig. S1b. The thermal noise and shot noise are probed at a high frequency range, where flicker noise is negligible (280-290 kHz), and their contributions are subtracted from the total measured noise. The obtained excess noise (Fig. S1c) is practically the flicker noise component. The measured flicker noise can be fitted to Hooge's expression[2]: $S_f \sim 1/f^\alpha$ (Fig. S1d), revealing $\alpha \approx 1$, as exemplified in Fig. S1e. The excess noise (Fig. S1c) is integrated in the range of $10^3$-$10^4$ Hz, in order to study the flicker noise as a function of conductance (Fig. S1f).

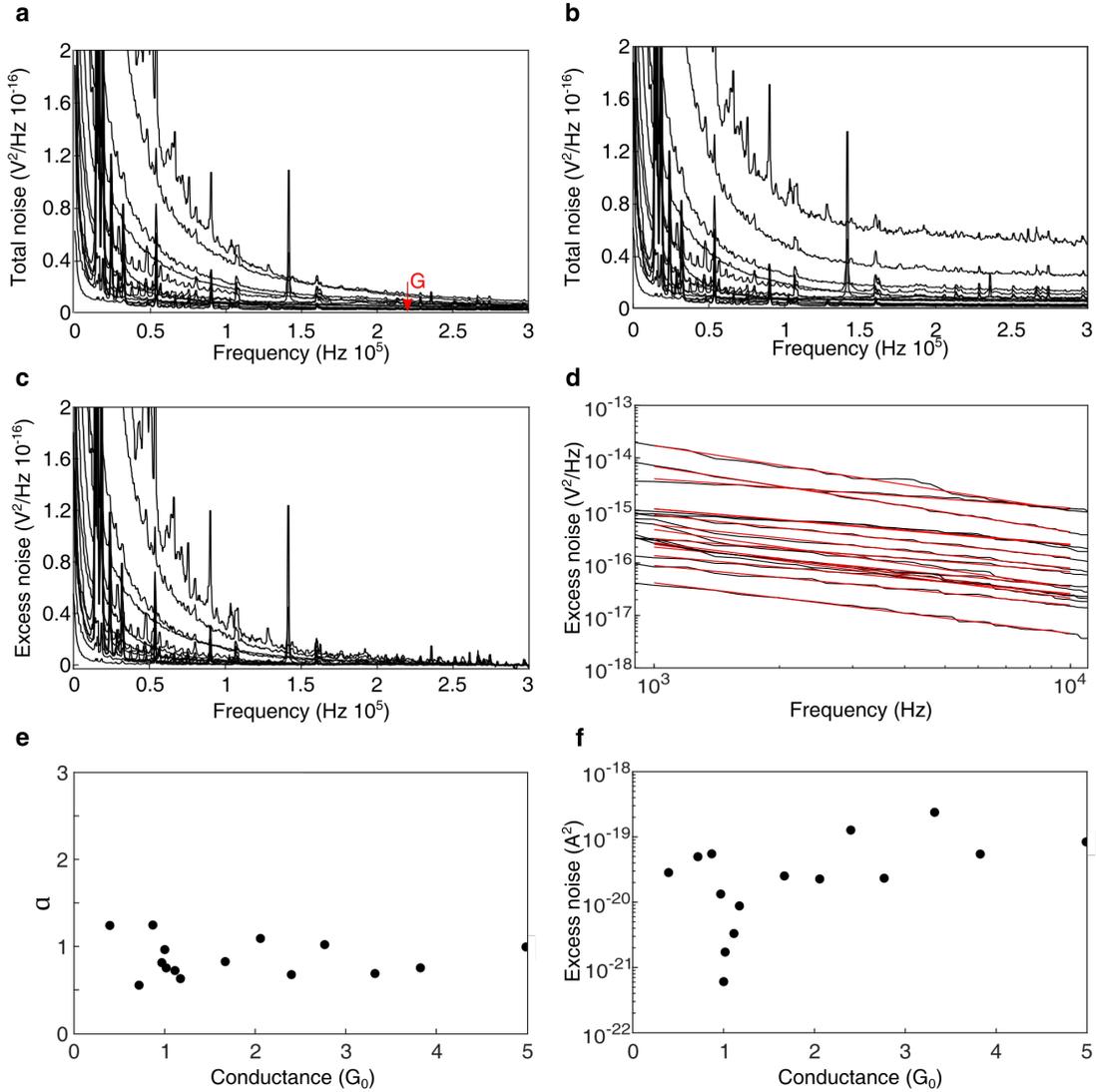

Figure S1: (a) Total noise as a function of frequency at 5mV, for $G = 0.39$-$4.99 \pm 0.01$ $G_0$. The top spectrum corresponds to the junction with the lowest conductance. The noise is suppressed by low-pass RC filtering owing to the capacitance of the setup, as well as the finite sample and wire resistances. (b) The data presented in (a) minus the setup voltage noise, and corrected by an RC transfer function followed by subtraction of the amplifier input current noise. (c) The excess noise that remains in the data presented in (b) after the average value in the frequency range of 280-290 kHz was subtracted from each spectrum. The received excess noise is practically the flicker noise contribution. (d) The data presented in (c) in a log scale view (black line), fitted to the classical flicker noise model: $S_f = S/f^{\alpha}$ (red line). (e) $\alpha$ as a function of conductance obtained from the fitted data presented in (d). (f) Excess noise as a function of conductance obtained by summing the noise contribution presented in (d) in the frequency range of 1-10 kHz. The data were converted from voltage power units ($V^2$) to current power units ($A^2$) by dividing each value by the square of the corresponding junction's resistance. The data error range is comparable or smaller than the diameter of the symbols.

### D. Flicker noise dependence on voltage

Figure S2 presents the voltage dependence of the extracted prefactors $S_{min}$ and $S_{max}$, revealing the typical quadratic dependence of flicker noise on voltage in the examined 5 mV range. The voltage experienced by the junction is an outcome of the applied current bias and the junction's resistance. Note that due to voltage-induced changes in the interference pattern of the scattered electronic wave functions in the vicinity of the junction[3,4], the simple quadratic dependence should not necessarily be preserved for a larger voltage range.

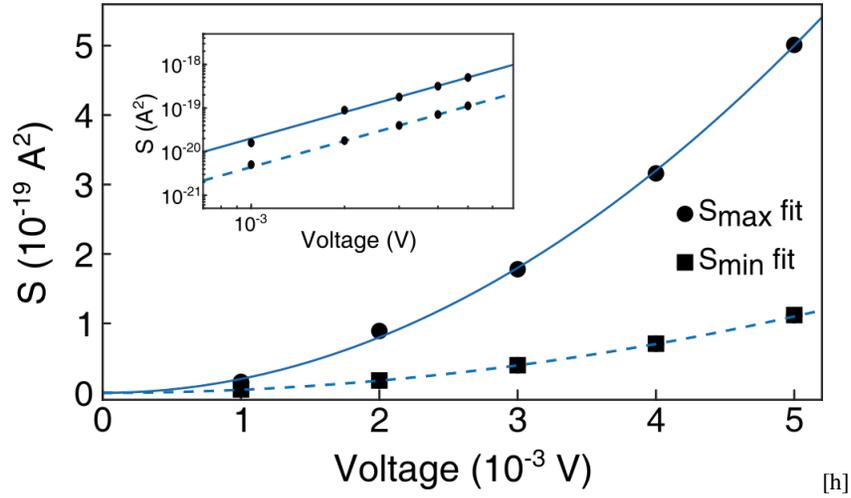

Figure S2: Prefectors $S_{min}$ and $S_{max}$ as a function of the induced voltage across the examined junctions. Inset: the same data in a log-log scale. Quadratic fits (blue) of $S = 2.0 \times 10^{-14} \pm 0.1 \times 10^{-14} V^2$ with an R-square value of 0.996 for $S_{max}$ and $S = 4.4 \times 10^{-15} \pm 0.0 \times 10^{-15} V^2$ with an R-square value of 0.999 for $S_{min}$, indicate a clear dependence of the prefactors on $V^2$.

### E. Deviation from a single channel near 1 $G_0$

The semitransparent yellow region in Fig. S3 is given by assuming two conduction channels with equal transmission probabilities. This region captures most of the upward deviation of the data point spread as the conductance approaches 1 $G_0$, indicating a significant contribution from a second conductance channel in this conductance region.

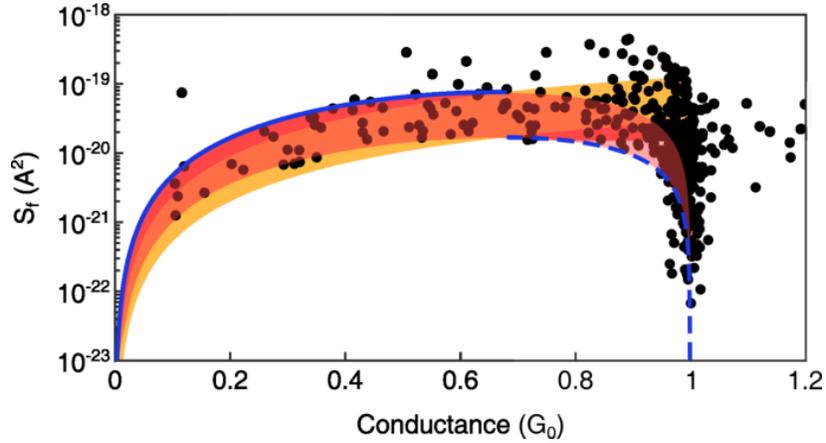

Figure S3: Flicker noise integrated in the range $10^3$-$10^4$ Hz as a function of conductance. Each data point (black) is measured for a different Au/hydrogen junction realization. The upper blue curve is a fit of $S_f = S_{max}\tau_1^2(1 - \tau_1)$ between 0 to 2/3 $G_0$ and the lower dashed blue curve is a fit to $S_f = S_{min}\tau_1^2(1 - \tau_1)$ between 2/3 and 1 $G_0$. The semitransparent red (yellow) region describes the expected flicker noise for a single (two equal) transmission channels, with $S$ between $S_{max}$ and $S_{min}$. The data error range is comparable or smaller than the diameter of the symbols.

# PART II: THEORY

## S2. SCATTERING THEORY

We focus on low-frequency electronic noise in atomic-scale junctions. Since the constriction is of an atomic size and the temperature is low, we assume that the time between inelastic scattering events of electrons by phonons is long relative to the elastic scattering timescale of electrons by e.g. defects. This understanding allows us to treat electrons as quantum coherent waves, and we use the Landauer-Buttiker formalism to describe their transport between metallic bulks. The elastic scattering of the electron wave critically depends on the arrangement of scatterers, such as defects and impurities in the conductor at the interface region close to the atomic constriction. We separate the junction into three regions:

(I) **Central region: atomic contact (C).** The central region of the junction is of an atomic dimensions and it supports ballistic transport. The width of this constriction is much smaller than the electrons' coherence length.

(II) **Interface zones.** To the left and right of the atomic constriction region, we have two interface regions with defects arising from lattice imperfections and impurities. Defects in these region are dynamic, resulting in the physics of the flicker noise, as we describe in this work. While the interface region is part of the macroscopic metal, the region considered lies within the coherent mean free path. Therefore, we describe scattering processes within the interface region using coherent scattering approach.

(III) **Metallic bulk.** Away from the two interface zones, beyond the phase-coherent and elastic mean free path, we treat the rest of the metal as an ideal lead (in the sense of noninteracting electron reservoir) and assume that the bulk can be described by a grand canonical ensemble at a given temperature and Fermi level.

While there are obviously defects in this region, and these defects may be slowly-moving as well, since this region lies outside the phase-coherent zone, this dynamics does not relate to the observed flicker noise and it affects timescale and the dynamics of local equilibration processes in the bulk.

Within the framework of scattering theory, the ballistic contact (I) and two interface zones (II) are treated as barriers characterized by a set of transmission and reflection amplitudes. To set the notation, we assume that there are $N$ channels; electrons in different channels have different transverse energies. For the atomic-scale region, we denote the $N \times N$ matrices of transmission amplitude $\boldsymbol{t}_C$ and reflection amplitude $\boldsymbol{r}_C$ for incoming electrons from the left side, and transmission amplitude $\boldsymbol{t}'_C$ and reflection amplitude $\boldsymbol{r}'_C$ for incoming electrons from the right side. Similarly, we introduce electron transmission amplitude matrices $\boldsymbol{t}_v$ and $\boldsymbol{t}'_v$, and reflection amplitude matrices, $\boldsymbol{a}_v$ and $\boldsymbol{a}'_v$ for the interface regions in the $v = L, R$ lead when coming from the left and right respectively.

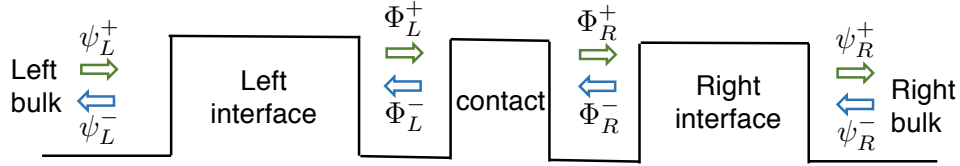

Figure S4: Schematic of scattering description of atomic junctions. $\psi_v^{\pm}$ and $\Phi_v^{\pm}$ with $v = L, R$ denote electron wave functions in $v$ lead with direction indicated by $\pm$.

We now derive, following[4], the total transmission probability in the junction for the multichannel case. For simplicity, hereafter we do not explicitly indicate the time and energy dependence of transmission and reflection amplitudes. We start from the (unitary) scattering matrix, which transforms incoming waves to outgoing ones. For the atomic-scale contact region, using the notations introduced in Fig. S4, we have

$$\begin{pmatrix} \Phi_L^- \\ \Phi_R^+ \end{pmatrix} = \boldsymbol{\mathcal{S}}_C \cdot \begin{pmatrix} \Phi_L^+ \\ \Phi_R^- \end{pmatrix} \tag{S1}$$

with the transmission and reflection amplitudes,

$$\boldsymbol{\mathcal{S}}_C = \begin{pmatrix} \boldsymbol{r}_C & \boldsymbol{t}'_C \\ \boldsymbol{t}_C & \boldsymbol{r}'_C \end{pmatrix} \tag{S2}$$

We define the transfer matrix $\boldsymbol{\mathcal{M}}_C$ for the atomic (C) region as

$$\begin{pmatrix} \Phi_R^+ \\ \Phi_R^- \end{pmatrix} = \boldsymbol{\mathcal{M}}_C \cdot \begin{pmatrix} \Phi_L^+ \\ \Phi_L^- \end{pmatrix}$$
$$= \begin{pmatrix} \boldsymbol{t}_C^{\dagger,-1} & \boldsymbol{r}'_C \boldsymbol{t}'^{-1}_C \\ -\boldsymbol{t}'^{-1}_C \boldsymbol{r}_C & \boldsymbol{t}'^{-1}_C \end{pmatrix} \begin{pmatrix} \Phi_L^+ \\ \Phi_L^- \end{pmatrix} \tag{S3}$$

In the second line, we expressed the transfer matrix $\boldsymbol{\mathcal{M}}_C$ in terms of transmission and reflection amplitudes. Similarly, for the left (L) and right (R) interface regions, we have ($v = L, R$)

$$\boldsymbol{\mathcal{M}}_v = \begin{pmatrix} \boldsymbol{t}_v^{\dagger, -1} & \boldsymbol{a}_v' \boldsymbol{t}_v'^{-1} \\ -\boldsymbol{t}_v'^{-1} \boldsymbol{a}_v & \boldsymbol{t}_v'^{-1} \end{pmatrix} \tag{S4}$$

The overall transfer matrix $\boldsymbol{\mathcal{M}}_{\mathrm{tot}}$ is obtained as $\boldsymbol{\mathcal{M}}_{\mathrm{tot}} = \boldsymbol{\mathcal{M}}_R \cdot \boldsymbol{\mathcal{M}}_C \cdot \boldsymbol{\mathcal{M}}_L$. Formally, we express $\boldsymbol{\mathcal{M}}_{\mathrm{tot}}$ as

$$\boldsymbol{\mathcal{M}}_{\mathrm{tot}} = \begin{pmatrix} \boldsymbol{t}_{\mathrm{tot}}^{\dagger, -1} & \boldsymbol{r}_{\mathrm{tot}}' \boldsymbol{t}_{\mathrm{tot}}'^{-1} \\ -\boldsymbol{t}_{\mathrm{tot}}'^{-1} \boldsymbol{r}_{\mathrm{tot}} & \boldsymbol{t}_{\mathrm{tot}}'^{-1} \end{pmatrix}, \tag{S5}$$

but using the detailed matrix forms for $\boldsymbol{\mathcal{M}}_C$ and $\boldsymbol{\mathcal{M}}_v$, we identify the inverse of the total transmission amplitude, which is just the matrix element $\boldsymbol{\mathcal{M}}_{\mathrm{tot}}^{22}$,

$$\begin{aligned} \boldsymbol{t}_{\mathrm{tot}}'^{-1} &= [\boldsymbol{\mathcal{M}}_R \cdot \boldsymbol{\mathcal{M}}_C \cdot \boldsymbol{\mathcal{M}}_L]_{22} \\ &= \boldsymbol{t}_R'^{-1} \left[ \boldsymbol{t}_C'^{-1} - \boldsymbol{a}_R \boldsymbol{t}_C^{\dagger, -1} \boldsymbol{a}_L' - \boldsymbol{t}_C'^{-1} \boldsymbol{r}_C \boldsymbol{a}_L' - \boldsymbol{a}_R \boldsymbol{r}_C' \boldsymbol{t}_C'^{-1} \right] \boldsymbol{t}_L'^{-1}. \end{aligned} \tag{S6}$$

Before proceeding, we make several approximations:

(i) Matrices of transmission and reflection amplitudes of the atomic-scale contact can be approximated as diagonals due to vanishing transitions between different channels across the contact. We can thus introduce the transmission probabilities $\{\tau_i\}_{i=1,2,\cdots,N}$ of the atomic-scale contact based on the diagonal matrix elements: $\tau_i \equiv |t_{C,ii}|^2 = |t'_{C,ii}|^2$, and $|r_{C,ii}|^2 = |r'_{C,ii}|^2 = 1 - \tau_i$.

(ii) Due to the phase-coherent nature of electron transport in atomic junctions at low temperatures, it is expected that elements of reflection amplitude matrices $\boldsymbol{a}_v$ and $\boldsymbol{a}_v'$ are small, and can be treated as perturbations when evaluating the total transmission probability.

(iii) As defects in general follow slow kinetics, the reflection amplitude matrices $\boldsymbol{a}_v(t)$ and $\boldsymbol{a}_v'(t)$, and consequently $\boldsymbol{t}_{\mathrm{tot}}'(t)$, should be time dependent over long timescales. We assume that the time dependence of the atomic arrangement at the contact (region I) is slower, and thus ignore it at this point.

We proceed and identify the transmission probability and organize it as

$$\begin{aligned} \mathcal{T} &\equiv \mathrm{Tr}[\boldsymbol{t}_{\mathrm{tot}} \boldsymbol{t}_{\mathrm{tot}}^{\dagger}] = \mathrm{Tr}[\boldsymbol{t}_{\mathrm{tot}}' \boldsymbol{t}_{\mathrm{tot}}'^{\dagger}] \\ &= \mathrm{Tr}[(\boldsymbol{t}_{\mathrm{tot}}'^{-1})^{\dagger} \boldsymbol{t}_{\mathrm{tot}}'^{-1}]^{-1}, \end{aligned} \tag{S7}$$

where we assumed time-reversal symmetry. The transmission $\boldsymbol{t}_{\mathrm{tot}}'(t)$ can be obtained from Eq. (S6).

Next, we utilize the fact that reflection amplitudes $\boldsymbol{a}_v$ and $\boldsymbol{a}_v'$ can be treated as perturbations in phase-coherent atomic junctions at low temperatures, such that the transmission amplitudes $\boldsymbol{t}_v \simeq \boldsymbol{I}$ and $\boldsymbol{t}_v' \simeq \boldsymbol{I}$ with $\boldsymbol{I}$ an $N \times N$ identity matrix. We then expand the following expression and keep terms up to the lowest order of $\boldsymbol{a}_v$ and $\boldsymbol{a}_v'$:

$$\begin{aligned} (\boldsymbol{t}_{\mathrm{tot}}'^{-1})^{\dagger} \boldsymbol{t}_{\mathrm{tot}}'^{-1} &\approx \left[ \boldsymbol{t}_C'^{-1} - \boldsymbol{t}_C'^{-1} \boldsymbol{r}_C \boldsymbol{a}_L' - \boldsymbol{a}_R \boldsymbol{r}_C' \boldsymbol{t}_C'^{-1} \right]^{\dagger} \left[ \boldsymbol{t}_C'^{-1} - \boldsymbol{t}_C'^{-1} \boldsymbol{r}_C \boldsymbol{a}_L' - \boldsymbol{a}_R \boldsymbol{r}_C' \boldsymbol{t}_C'^{-1} \right] \\ &\approx (\boldsymbol{t}_C' \boldsymbol{t}_C'^{\dagger})^{-1} - (\boldsymbol{t}_C'^{-1})^{\dagger} \left[ \boldsymbol{t}_C'^{-1} \boldsymbol{r}_C \boldsymbol{a}_L' + \boldsymbol{a}_R \boldsymbol{r}_C' \boldsymbol{t}_C'^{-1} \right] \\ &\quad - \left[ \boldsymbol{t}_C'^{-1} \boldsymbol{r}_C \boldsymbol{a}_L' + \boldsymbol{a}_R \boldsymbol{r}_C' \boldsymbol{t}_C'^{-1} \right]^{\dagger} \boldsymbol{t}_C'^{-1}. \end{aligned} \tag{S8}$$

Inserting the above expression into Eq. (S7), the total transmission function is obtained as

$$\begin{aligned} \mathcal{T} &\approx \sum_{i=1}^{N} \left[ \tau_i^{-1} - \tau_i^{-1} 2\mathrm{Re} \left( r_{C,ii} a_{L,ii}' + a_{R,ii} r_{C,ii}' \right) \right]^{-1} \\ &\approx \sum_{i=1}^{N} \tau_i \left[ 1 + 2\mathrm{Re} \left( r_{C,ii} a_{L,ii}' + a_{R,ii} r_{C,ii}' \right) \right]. \end{aligned} \tag{S9}$$

Here, 'Re' takes the real part. To obtain the first line, we assumed that the transmission and reflection processes between different channels are negligible, hence the corresponding amplitude matrices can be treated as diagonals. As a reminder, the transmission probability depends on time and the energy of incoming electrons, $\mathcal{T}(\epsilon, t)$.

## S3. DERIVATION OF FLICKER NOISE

### A. Current fluctuations

As dynamics of defects occurs on a time scale that is much longer than that of electron tunneling through the contact, we can still adopt the Landauer-Büttiker expression for describing charge current in the junction,

$$I(t) = \frac{2e}{h} \int_{-\infty}^{+\infty} d\epsilon \, \mathcal{T}(\epsilon, t) [f_L(\epsilon) - f_R(\epsilon)]. \tag{S10}$$

Here, $f_v(\epsilon) = \{\exp[(\epsilon - \mu_v)/k_B T] + 1\}^{-1}$ denotes the Fermi-Dirac distribution for the $v$ lead with Fermi level $\mu_v$ and a constant temperature $T$. The transmission function is give by Eq. (S9). The transmission probability of electrons in the junction is changing in time due to the slow dynamics of defects at the interface zones, specifically altering the return amplitudes $a_\nu$ and $a'_\nu$. Furthermore, the transmission probability depends on energy of the incoming electrons, $\epsilon$. However, for atomic junctions at low voltage the energy dependence of the transmission probability boils down to the behavior close to the Fermi energy, as we show below.

Explicitly noting the energy and time dependence we write

$$
\begin{aligned}
\mathcal{T}(\epsilon, t) &\simeq \sum_{i=1}^{N} \tau_i(\epsilon) \left[ 1 + 2\mathrm{Re}\left( \boldsymbol{r}_{C,ii}(\epsilon) \boldsymbol{a}'_{L,ii}(\epsilon, t) + \boldsymbol{a}_{R,ii}(\epsilon, t) \boldsymbol{r}'_{C,ii}(\epsilon) \right) \right] \\
&= \langle \mathcal{T}(\epsilon) \rangle + \delta\mathcal{T}(\epsilon, t),
\end{aligned}
\tag{S11}
$$

identifying

$$
\begin{aligned}
\tau(\epsilon) &\equiv \langle \mathcal{T}(\epsilon) \rangle = \sum_{i=1}^{N} \tau_i(\epsilon), \\
\delta\mathcal{T}(\epsilon, t) &= \sum_{i=1}^{N} 2\tau_i(\epsilon) \mathrm{Re}\left( \boldsymbol{r}_{C,ii}(\epsilon) \boldsymbol{a}'_{L,ii}(\epsilon, t) + \boldsymbol{a}_{R,ii}(\epsilon, t) \boldsymbol{r}'_{C,ii}(\epsilon) \right).
\end{aligned}
\tag{S12}
$$

Inserting the above expressions into Eq. (S10), we have $I(t) = \langle I \rangle + \delta I(t)$ with

$$
\delta I(t) = \frac{2e}{h} \int_{-\infty}^{+\infty} d\epsilon \, \delta\mathcal{T}(\epsilon, t) [f_L(\epsilon) - f_R(\epsilon)],
\tag{S13}
$$

which is just the current fluctuations induced by defects in the interface regions. Particularly, from the expression of $\delta\mathcal{T}(\epsilon, t)$ in Eq. (S12), we see that return processes in which electrons from the atomic region are reflected back to that region by defects play a dominant role in inducing current fluctuations.

Denoting $\mu_L - \mu_R = eV$ with $V$ the applied voltage bias, in the limits of low temperature and for small applied voltage bias we can simplify the integrals in Eq. (S13), e.g.

$$
\begin{aligned}
\int_{-\infty}^{+\infty} d\epsilon \tau_i(\epsilon) \boldsymbol{r}_{C,ii}(\epsilon) \boldsymbol{a}_{v,ii}(\epsilon, t) [f_L(\epsilon) - f_R(\epsilon)] &\approx \int_{-\infty}^{+\infty} d\epsilon \tau_i(\epsilon) \boldsymbol{r}_{C,ii}(\epsilon) \boldsymbol{a}_{v,ii}(\epsilon, t) \left( \frac{\partial f}{\partial \mu} \right) (\mu_L - \mu_R) \\
&= eV \int_{-\infty}^{+\infty} d\epsilon \tau_i(\epsilon) \boldsymbol{r}_{C,ii}(\epsilon) \boldsymbol{a}_{v,ii}(\epsilon, t) \left( -\left. \frac{\partial f}{\partial \epsilon} \right|_{\mu=\epsilon_F} \right) \\
&\approx eV \tau_i(\epsilon_F) \boldsymbol{r}_{C,ii}(\epsilon_F) \boldsymbol{a}_{v,ii}(\epsilon_F, t).
\end{aligned}
\tag{S14}
$$

Here, $f(\epsilon) = \{\exp[(\epsilon - \mu)/k_B T] + 1\}^{-1}$ and $\epsilon_F$ denotes the Fermi energy. Below we omit the energy dependence of these amplitudes, since it is trivial.

## B. Power spectrum of current fluctuations

Flicker noise corresponds to the low frequency behavior of the power spectrum of the fluctuating current. Before proceeding, we note that the dynamics of $\delta I(t)$ depends on the details of the samples, such as the number of defects, and thus may vary between samples. One may adopt a generic sample spectrum to analyze the power spectrum of $\delta I(t)$ recorded in a time interval $(0, t_m)$ (see e.g., Refs.[5,6]). In what follows, our focus is on the dependence of the flicker noise on the channel distribution, rather than on the details of the frequency dependence. We denote the flicker noise by

$$
S_f(\omega) = \lim_{t_m \to \infty} \frac{1}{t_m} \left| \int_0^{t_m} \delta I(t) e^{i\omega t} dt \right|^2,
\tag{S15}
$$

investigated in the limit of a long measurement time $t_m$. Note that the above definition can be applied to any random process, be it ergodic or non-ergodic, stationary or non-stationary. If the random process $\delta I(t)$ is stationary in the long time limit, the above spectrum reduces to the fundamental Wiener-Khinchin theorem, which relates the power spectrum to the correlation function[7].

Using Eq. (S13) together with Eqs. (S12) and (S14), we have

$$
\delta I(t) \simeq \sum_i 2\tau_i G_0 V \mathrm{Re}\left[ \boldsymbol{r}_{C,ii} \boldsymbol{a}'_{L,ii}(t) + \boldsymbol{a}_{R,ii}(t) \boldsymbol{r}'_{C,ii} \right].
\tag{S16}
$$

We assume that only reflection amplitudes due to defects at the interface regions vary in time. Inserting the above expression into the spectrum definition Eq. (S15), we find

$$
\begin{aligned}
S_f(\omega) &\simeq 4G_0^2 V^2 \lim_{t_m \to \infty} \frac{1}{t_m} \left| \sum_i \tau_i \mathrm{Re}\left[ \boldsymbol{r}_{C,ii} \boldsymbol{a}'_{L,ii}(\omega) + \boldsymbol{a}_{R,ii}(\omega) \boldsymbol{r}'_{C,ii} \right] \right|^2 \\
&\simeq 4G_0^2 V^2 \sum_i \tau_i^2 \lim_{t_m \to \infty} \frac{1}{t_m} \left( \mathrm{Re}\left[ \boldsymbol{r}_{C,ii} \boldsymbol{a}'_{L,ii}(\omega) + \boldsymbol{a}_{R,ii}(\omega) \boldsymbol{r}'_{C,ii} \right] \right)^2 \\
&= 4G_0^2 V^2 \sum_i \tau_i^2 \lim_{t_m \to \infty} \frac{1}{t_m} \frac{1}{4} \left( \boldsymbol{r}_{C,ii} \boldsymbol{a}'_{L,ii}(\omega) + \boldsymbol{a}_{R,ii}(\omega) \boldsymbol{r}'_{C,ii} + \mathrm{c.c.} \right)^2 \\
&\approx 4G_0^2 V^2 \sum_i \tau_i^2 \lim_{t_m \to \infty} \frac{1}{t_m} \frac{1}{2}(1-\tau_i) \sum_{v=L,R} |\boldsymbol{a}_{v,ii}(\omega)|^2 \\
&= 2G_0^2 V^2 \sum_i \tau_i^2 (1-\tau_i) \sum_{v=L,R} \Phi_{v,ii}(\omega).
\end{aligned}
\tag{S17}
$$

Here, $G_0 = 2e^2/h$ is the conductance quantum and 'c.c.' stands for complex conjugate in this scenario. We also defined $\boldsymbol{a}_{v,ii}(\omega) \equiv \int_0^{t_m} \boldsymbol{a}_{v,ii}(t) e^{i\omega t} dt$ and $\Phi_{v,ii}(\omega) = \lim_{t_m \to \infty} \left| \int_0^{t_m} \boldsymbol{a}_{v,ii}(\epsilon_F, t) e^{i\omega t} dt \right|^2/t_m$. To obtain the second line, we neglected correlations between different channels, which is consistent with the derivation of total transmission function. In obtaining the fourth line, we just kept terms with nonzero contributions. The transmission probability $\tau_i$ is evaluated at the Fermi energy. We assume that electrons in different channels are affected by the same defect configuration, thus the spectrum $\Phi_{v,ii}(\omega)$ is independent of the channel index $i$ such that

$$
S_f(\omega) \simeq 2G_0^2 V^2 \sum_{v=L,R} \Phi_v(\omega) \sum_{i=1}^N \tau_i^2(1-\tau_i),
\tag{S18}
$$

which is Eq. (1) in the main text. From the above expression, we conclude that the low-frequency behavior of $S_f(\omega)$ is fully determined by that of $\Phi_v(\omega)$. Recall however that Eq. (S17) represents the lowest order contribution of dynamic scatterers to the power spectrum (Eq. (S16) represents their lowest order contribution to the current fluctuation).

Eq. (S18) connects between the measured flicker noise and the microscopic picture of transmission channels. When the flicker noise Eq. (S18) is used in conjunction with measurements of the electronic conductance, $G = \sum_i \tau_i$ and the Fano factor emerging from Shot noise, $F = \sum_i \tau_i(1-\tau_i)$, one can more accurately identify the contribution of different channels to the total transmission process. This quadratic dependence of $S_f$ on voltage identifies resistance fluctuations as the source of the $1/f$ noise.

### C. Generalization to dynamical atomic contact

In a more general setting, we allow time-dependent transmission and reflection amplitudes for the ballistic contact region such that Eq. (S16) becomes

$$
\delta I(t) \simeq \sum_i 2\tau_i(t) G_0 V \mathrm{Re}\left[ \boldsymbol{r}_{C,ii} \boldsymbol{a}'_{L,ii}(\epsilon_F, t) + \boldsymbol{a}_{R,ii}(\epsilon_F, t) \boldsymbol{r}'_{C,ii}(t) \right].
\tag{S19}
$$

We follow the same procedure that leads to Eq. (S17) and find

$$
\begin{aligned}
S_f(\omega) &\simeq 4G_0^2 V^2 \lim_{t_m \to \infty} \frac{1}{t_m} \left| \sum_i \mathrm{Re}\left[ \left( \tau_i \boldsymbol{r}_{C,ii} \boldsymbol{a}'_{L,ii} \right)_\omega + \left( \tau_i \boldsymbol{a}_{R,ii} \boldsymbol{r}'_{C,ii} \right)_\omega \right] \right|^2 \\
&\simeq 4G_0^2 V^2 \sum_i \lim_{t_m \to \infty} \frac{1}{t_m} \left( \mathrm{Re}\left[ \left( \tau_i \boldsymbol{r}_{C,ii} \boldsymbol{a}'_{L,ii} \right)_\omega + \left( \tau_i \boldsymbol{a}_{R,ii} \boldsymbol{r}'_{C,ii} \right)_\omega \right] \right)^2 \\
&\approx 2G_0^2 V^2 \sum_i \lim_{t_m \to \infty} \frac{1}{t_m} \left[ \left| \left( \tau_i \boldsymbol{r}_{C,ii} \boldsymbol{a}'_{L,ii} \right)_\omega \right|^2 + \left| \left( \tau_i \boldsymbol{a}_{R,ii} \boldsymbol{r}'_{C,ii} \right)_\omega \right|^2 \right] \\
&\equiv 2G_0^2 V^2 \sum_{i,v} \tilde{\Phi}_{v,ii}(\omega).
\end{aligned}
\tag{S20}
$$

Here, $A_\omega \equiv \int_0^{t_m} A(t) e^{i\omega t} dt$ for an arbitrary time-dependent function $A(t)$. In the long time limit the power spectrum can be written as a convolution of functions, of the interface zone and the atomic contact. For example,

$$
\begin{aligned}
\tilde{\Phi}_{L,ii}(\omega) &= \frac{1}{t_m} \left| \int_0^{t_m} \tau_i(t) \boldsymbol{r}_{C,ii}(t) \boldsymbol{a}'_{L,ii}(t) e^{i\omega t} dt \right|^2 \\
&\approx \frac{1}{(2\pi)^2 4 t_m} \left| \int_{-\infty}^{\infty} d\omega_1 F_{C,i}(\omega_1) \boldsymbol{a}_{L,ii}{}'(\omega - \omega_1) \right|^2,
\end{aligned}
\tag{S21}
$$

where $F_{C,i}(\omega) = \int_{-t_m}^{t_m} dt\, e^{i\omega t} \tau_i(t) \boldsymbol{r}_{C,ii}(t)$ and $\boldsymbol{a}_{L,ii}(\omega) = \int_{-t_m}^{t_m} dt\, e^{i\omega t} \boldsymbol{a}'_{L,ii}(t)$. Assuming time scale separation between the two processes, that is, $\boldsymbol{a}_{L,ii}(\omega - \omega_1) \approx \boldsymbol{a}_{L,ii}(\omega)$ with the atomic constriction static on the time scale of the dynamics of impurities in the interface zone, we recover Eq. (S17).

### D. Relation to conductance fluctuations

The functional form of Eq. (S18) with transmission was observed in measurements of fluctuations in the conductance with bias voltage in atomic-size constrictions, $\langle \delta G(V_1) \delta G(V_2) \rangle$, see Refs.[3,4]. Here, $\delta G$ is the fluctuations of the conductance from the mean, $\delta G = G - \langle G \rangle$. $V$ is the voltage bias, and the average is done over the ensemble of defect configurations in the "diffusive bank", the metallic region beyond the constriction[4]. We now detail on the mathematical and physical relationship between conductance fluctuations in ballistic conductors and the quantum flicker noise, which we study here.

Conductance fluctuations in atomic-size gold contacts were investigated experimentally in Refs.[3,4] testing different metals, further developing an accompanying quantum scattering formalism based on the Landauer-Buttiker theory to describe the effect. It was argued that the fluctuations in the conductance as a function of bias voltage are the result of interference effects between different pathways, with electron waves either directly transmitted between metals, or first scattered by defects at the diffusive banks. Physically, the origin of conductance fluctuations is that a change in voltage modifies the kinetic energy of electrons, thus it modulates the phase of interfering electrons, showing up as conductance fluctuations. It was shown in Refs.[3,4] that $\left\langle \left( \frac{\partial \delta G}{\partial V} \right)^2 \right\rangle \propto \sum_i \tau_i^2 (1 - \tau_i)$, and similarly, $\langle \delta G(V_1) \delta G(V_2) \rangle \propto \sum_i \tau_i^2 (1 - \tau_i)$. In the latter expression, one studies the autocorrelation function of the conductance at different voltage bias. In contrast, flicker noise describe fluctuations in the current noise over time, $\langle \delta I(t) \delta I(t + t') \rangle = \frac{1}{T} \int_{-T/2}^{T/2} dt\, \delta I(t) \delta I(t + t')$ (for simplicity, assuming an ergodic process and the validity Wiener-Khinchin theorem), which after Fourier transform results in the power spectrum.

Scatterers in the region proximal to the atomic contact play a decisive role in both effects in atomic-scale junctions, conductance fluctuations and the quantum flicker noise. In the effect of conductance fluctuations[3,4], we do not account for the dynamics of scatterers, but study the role of static scatterers in interference effects, while modifying voltage. Looking at frequency-dependent noise, we probe the time autocorrelation function of the current; this function reveals (to some extent) the motion of scatterers. Both conductance fluctuations and quantum flicker noise in atomic-size junctions reveal the critical role of scattering effects beyond the atomic constriction in dictating interference effects of electrons through the junction.

### E. Beyond the noninteracting electron model

Electron-phonon couplings leave signatures in the shot noise, providing e.g. information on the local phonon population and the lattice temperature[8,9]. To asses the role of electron-phonon couplings (here we consider phonon modes of the central region) in flicker noise, we assume the following simple decomposition of the transmission probability, $\tau \approx \tau_{el} + \tau_{el-ph}$, where the first term does not depend on electron-phonon couplings, which are affecting only the second term. Such decompositions can be derived using perturbation theory assuming weak system-bath couplings. Assuming that electron-phonon coupling do not mix transmission channels, the derivation of the flicker noise follows as described above resulting in $S_f \propto \sum_i \tau_i^2 (1 - \tau_i)$. Focusing on the single channel case we write

$$
\begin{aligned}
S_f &\propto \tau^2 (1 - \tau) \\
&= (\tau_{el} + \tau_{el-ph})^2 [1 - (\tau_{el} + \tau_{el-ph})] \\
&\approx \tau_{el}^2 (1 - \tau_{el}) + 2\tau_{el-ph} \tau_{el} (1 - \tau_{el}) - \tau_{el}^2 \tau_{el-ph} \\
&= \tau_{el}^2 (1 - \tau_{el}) + \tau_{el-ph} [2\tau_{el} - 3\tau_{el}^2].
\end{aligned}
\tag{S22}
$$

Therefore, corrections to the purely coherent-electronic flicker noise would scale as $\tau_{el-ph}$. These corrections due to electron-phonon scatterings in the central region would increase the flicker noise if $\tau_{el} < 2/3$; for larger transmission constant, $\tau_{el} > 2/3$, vibration-mode scattering would suppress the flicker noise.

## S4.   BOUNDS ON THE CURRENT POWER SPECTRUM

### A.   Upper bound

Flicker noise can be written as $S_f(\omega) = S(\omega) \times \sum_{i=1}^{N} \tau_i^2(1-\tau_i)$, see Eq. (S18), with the prefactor $S(\omega)$ that per our assumptions, is independent of the channel index and the transmission probabilities, $\{\tau_i\}$. Here, we ask ourselves the following questions: Given a certain total conductance $\tau = \sum_{i=1}^{N} \tau_i$ due to the contribution of $N$ channels, what is the maximal value of the quantum flicker noise, $S_f(\omega)$? What are the corresponding contributions of each channel? To answer that, we consider the multi-variable function,

$$\mathcal{F}_N(\{\tau_i\}) \equiv \sum_{i=1}^{N} \tau_i^2(1-\tau_i), \tag{S23}$$

subjected to the constraint that the total transmission probability is fixed to a given value,

$$\tau = \sum_{i=1}^{N} \tau_i. \tag{S24}$$

If we define the function $g(\{\tau_i\}) = \sum_{i=1}^{N} \tau_i - \tau$, the constraint translates to $g(\{\tau_i\}) = 0$. We now define the Lagrange function,

$$\begin{aligned}\mathcal{L}(\{\tau_i\}, \lambda) &\equiv \mathcal{F}_N(\{\tau_i\}) - \lambda g(\{\tau_i\}) \\ &= \sum_{i=1}^{N} \tau_i^2(1-\tau_i) - \lambda\left(\sum_{i=1}^{N} \tau_i - \tau\right).\end{aligned} \tag{S25}$$

Here $\lambda$ is a Lagrange multiplier. The stationary point for the Lagrange function is obtained by solving the following $N$ equations,

$$\frac{\partial \mathcal{L}}{\partial \tau_j} = 2\tau_j - 3\tau_j^2 - \lambda = 0 \tag{S26}$$

along with the constraint, $\tau - \sum_i \tau_i = 0$. Since the above equation is identical for each $\tau_i$, we conclude that the stationary point requires $\tau_1 = \tau_2 = \cdots = \tau_N$, and based on the constraint, $\tau_i = \tau/N$ for each channel (Note that the quadratic equation Eq. (S26) has two solutions, but one of them is unphysical given that $0 \leq \tau_i \leq 1$).

It can be shown that this critical point is a local maximum if $\tau > N/3$ for $N$ channels. To see this, without loss of generality, we write

$$\mathcal{F}_N(\tau_1, \tau_2, ..., \tau_{N-1}; \tau_N) = \sum_{i=1}^{n-1} \tau_i^2(1-\tau_i) + \tau_N^2(1-\tau_N), \tag{S27}$$

where the transmission of the $N$th channel

$$\tau_N = \tau - \sum_{i=1}^{N-1} \tau_i, \tag{S28}$$

is written as a dependent variable, dependent on the transmissions of the first $N-1$ channels. The Hessian matrix can now be constructed by noting that

$$\frac{\partial^2}{\partial \tau_j \partial \tau_i} \mathcal{F}_N = 2 - 6\tau_N \; ; \; i \neq j \tag{S29}$$

and

$$\frac{\partial^2}{\partial \tau_i^2} \mathcal{F}_N = 2(2 - 6\tau_N) \tag{S30}$$

when evaluated at the extremal point, $\tau_i = \tau/N$. Explicitly, the Hessian matrix $H$ of dimensions $(n-1) \times (n-1)$ has the entries

$$H_{ij} = (2 - 6\tau_N)(1 + \delta_{ij}) = (2 - 6\tau_N)A_{ij}. \tag{S31}$$

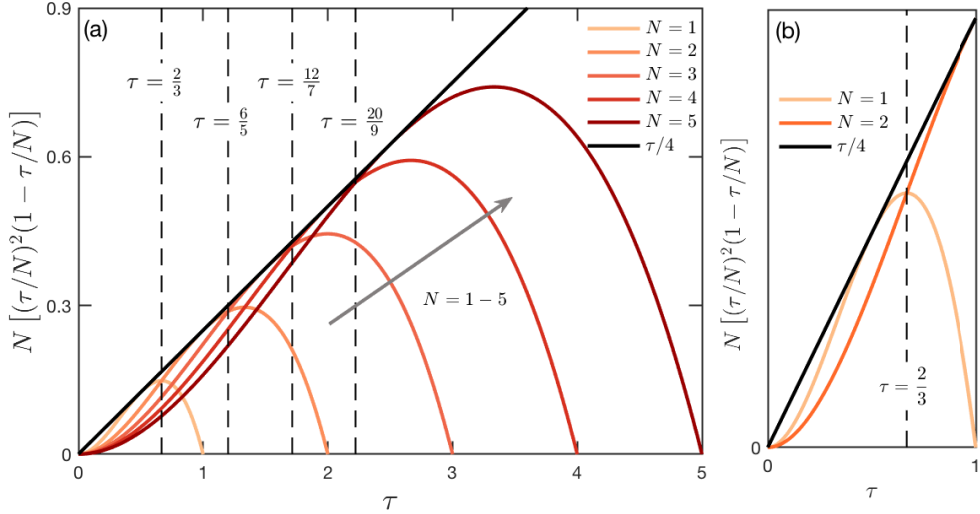

Figure S5: Upper bound of the transmission-dependent term in the flicker noise, $\mathcal{F}_N^{(max)}$. (a) We present Eq. (S35) as a function of $\tau$ for $N = 1 - 5$ (left to right), demonstrating that as we increase $\tau$, the noise is maximized by more channels. We further display the function $\tau/4$, which upper-bounds the flicker noise. (b) Zoom over the region $0 \le \tau \le 1$ displaying the relevant bounds $\mathcal{F}_1^{(max)}$ and $\mathcal{F}_2^{(max)}$, where the latter exceeds the former once $\tau > 2/3$. Vertical dashed line at $\tau$=2/3, 6/5, 12/7, 20/9 mark transmissions beyond which partitioning the transmission to $N + 1$ channels, rather than $N$ channels, maximizes the flicker noise, with $N = 2, 3, 4, 5$, respectively. For example, beyond $\tau = 2/3$ the flicker noise is maximized if two channels (equally) participate.

The matrix $A_{ij} = 1 + \delta_{ij}$ can be shown explicitly to be positive definite. We posit that the $N-1$ linearly independent eigenvectors of $A$ are given by

$$v_1 = \begin{bmatrix} 1 \\ 1 \\ \vdots \\ 1 \end{bmatrix} \tag{S32}$$

and the remaining $n - 2$ eigenvectors have the form

$$v_2 = \begin{bmatrix} 1 \\ -1 \\ 0 \\ \vdots \\ 0 \end{bmatrix}; \quad v_3 = \begin{bmatrix} 1 \\ 0 \\ -1 \\ \vdots \\ 0 \end{bmatrix}; \quad \ldots; \quad v_{N-1} = \begin{bmatrix} 1 \\ 0 \\ 0 \\ \vdots \\ -1 \end{bmatrix}; \tag{S33}$$

where $v_i$, for $1 < i \le N - 1$ has the first entry as 1 and the $i$-th entry as $-1$ and all other entries are zero. The reader may check that the vectors given above are indeed eigenvectors. The vector $v_1$ has an eigenvalue of $N$, $Av_1 = Nv_1$, whereas all other vectors have an eigenvalue of 1, $Av_i = v_i$ when $1 < i \le N - 1$. Since all eigenvalues of $A$ are positive, it is positive definite. Concluding, the Hessian Eq. (S31) is negative definite when

$$\tau_N = \frac{\tau}{N} > \frac{1}{3}, \tag{S34}$$

and the extremal point given $\tau_i = \tau/N$ maximizes the flicker noise. If the conductances instead satisfy $\tau_i = \tau_N < 1/3$, then the extremal point minimizes the flicker noise.

Based on this finding, assuming that $N$ channels contribute to the conductance and that the total conductance is large enough such that $\tau > N/3$ (which we will assume henceforth), $\mathcal{F}_N$ is upper bounded by the value

$$\mathcal{F}_N^{\max} = N \left[ \left( \frac{\tau}{N} \right)^2 (1 - \tau/N) \right], \tag{S35}$$

which is reached when $\tau_1 = \tau_2 = \tau_3 = \ldots = \tau_N = \tau/N$.

For example, if only a single channel is active, $\mathcal{F}_1^{\max} = \tau^2(1-\tau)$, and the maximum of this function with respect to the transmission is obtained at $\tau = 2/3$; $F_1^{\max}$ is upper bounded by the fraction $4/27$. If two channels contribute, the maximal noise is given by $\mathcal{F}_2^{\max} = \frac{\tau^2}{2}(1-\tau/2)$, obtained for $\tau = 4/3$. However, when the total conductance is such that $\tau < N/3$, the flicker noise is minimized at equal partition.

Next, we study the extremum of Eq. (S35) with respect to the number of channels. The derivative of the equi-partitioned flicker noise is

$$\frac{d\mathcal{F}_N^{\max}}{dN} = -\frac{\tau^2}{N^2} + \frac{2\tau^3}{N^3}, \tag{S36}$$

leading to a maximum in the flicker noise for $N^*$ channels with

$$N^* = 2\tau. \tag{S37}$$

Plugging this value in Eq. (S35) we conclude that for a given conductance ($\tau$) the transmission-dependent factor of the flicker noise is upper bounded by

$$\mathcal{F}(\tau) \leq \frac{\tau}{4}. \tag{S38}$$

This bound was achieved by maximizing the flicker noise with respect to channel distribution and with respect to the number of channels. Eq. (S38) is one central result of our work. For a given conductance $G = \sum_i \tau_i$, the level of flicker noise cannot exceed $G/4$.

Note that since

$$\frac{\partial^2}{\partial N^2}\mathcal{F}_N^{\max} = 2\frac{\tau^2}{N^3} - 6\frac{\tau^3}{N^4} = \frac{\tau^2}{N^3}\left(2 - \frac{6\tau}{N}\right) < 0 \text{ when } N = N^* = 2\tau, \tag{S39}$$

the second derivative of $\mathcal{F}$ with respect to $N$ has the same sign as the Hessian (S31). That is, $\mathcal{F}$ admits a maxima w.r.t $N$ whenever it admits a maxima w.r.t the transmissions. It may admit a minima w.r.t $N$ whenever $\mathcal{F}$ admits a minima w.r.t the transmissions. However, the extremal point $N^* = 2\tau$ precludes the admission of minima for the transmissions (since the Hessian is negative). This implies that $\mathcal{F}_N^{\min}$ in fact is also maximized w.r.t $N$ at $N = N^*$.

We now improve Eq. (S38) by upper-bounding the flicker noise *tightly*, albeit with a piecewise function. Let us ask first the following question: When is this inequality $\mathcal{F}_2^{\max} > \mathcal{F}_1^{\max}$ valid? It is immediate to find that

$$\frac{\tau^2}{2}(1-\tau/2) > \tau^2(1-\tau) \tag{S40}$$

once $\tau > 2/3$. Therefore (i) A single channel upper-bounds the noise between $1/3 \leq \tau \leq 2/3$. (ii) $\mathcal{F}_1^{\max}$ is maximized at $\tau = 2/3$. (ii) Beyond that, up to $\tau = 6/5$, the noise is maximized when two channels contribute.

More generally, let us consider two possible resolutions for the transmission with either $N$ or $M$ channels, with $N > M$. It is straightforward to prove that

$$\mathcal{F}_N^{\max} = N\frac{\tau^2}{N^2}(1-\tau/N) > M\frac{\tau^2}{M^2}(1-\tau/M) = \mathcal{F}_M^{\max} \tag{S41}$$

once

$$\tau > \frac{MN}{M+N}. \tag{S42}$$

Back to Eq. (S37), since $N^*$ is not necessarily an integer, we have to consider its ceiling $\bar{N}^*$ and floor, $\underline{N}^* = \bar{N}^* - 1$, which are the closest integers. Based on Eq. (S42), $\mathcal{F}_{\bar{N}^*}^{\max} > \mathcal{F}_{\bar{N}^*-1}^{\max}$ if

$$\tau > \frac{\bar{N}^*(\bar{N}^*-1)}{(2\bar{N}^*-1)}. \tag{S43}$$

For example, suppose that the total transmission is $\tau = 3/4$. In this case, $N^* = 3/2$, $\bar{N}^* = 2$ and $\bar{N}^* - 1 = 1$. Since $\tau = 3/4 > 2/3$ [condition (S43)] the flicker noise is maximized when two channels contribute. In contrast, for $\tau = 7/12$, $N^* = 7/6$, and $\bar{N}^* = 2$. In this case, we do not satisfy (S43), $\tau = 7/12 < 2/3$, therefore a single channel maximizes the flicker noise.

Altogether, a tight bound piece-wise function for the flicker noise is given by the following procedure: (i) Calculate $N^* = 2\tau$, its ceiling $\bar{N}^*$ and floor $\bar{N}^* - 1$. (ii) Test the inequality (S43). If it is satisfied, the maximal noise is given by $\mathcal{F}_{\bar{N}^*}^{\max}$. Else, it is determined by $\mathcal{F}_{\bar{N}^*-1}^{\max}$. These functions are calculated with Eq. (S35).

In Fig. S5, we present the function $\mathcal{F}_N^{\max}$ [Eq. (S35)], which is the maximal noise for a given transmission probability and assuming $N$ channels. By increasing the number of channels we demonstrate how the piecewise upper bound is constructed. Furthermore, we show that the simpler form, $\mathcal{F}^{\max} = \tau/4$, upper-bounds the flicker noise.

### B. Upper and lower bounds, $0 < \tau \leq 1$.

We now focus on the region between $0 < \tau \leq 1$. We will assume that at most two channels contribute in this region, which is a good approximation at low conductance.

First, according to Eq. (S40), a single channel upper-bounds the flicker noise for $1/3 < \tau \leq 2/3$. Next, we prove that for $2/3 < \tau < 1$, two channels with any partition always provide higher noise than a single channel. As a result, flicker noise arising from a single open channel would define the lower bound of flicker noise in this regime.

Consider the total transmission $\tau$ with two possible decompositions, to a single channel, or to two channels with $\tau_1$ and $\tau - \tau_1$ as transmission probabilities. We ask ourselves when does the following inequality for the flicker noise hold (for the nontrivial channel decomposition $\tau_1 \neq 0$),

$$\tau^2(1-\tau) \leq \tau_1^2(1-\tau_1) + (\tau-\tau_1)^2(1-\tau+\tau_1). \tag{S44}$$

After simple manipulations, we get

$$\tau_1(3\tau - 1) \leq \tau(3\tau - 2), \tag{S45}$$

which, for $\tau > 1/3$, reduces to

$$\tau_1 \leq \tau\left(\frac{3\tau - 2}{3\tau - 1}\right), \tag{S46}$$

Since $\tau_1$ must be smaller than $\tau$, both positive, this inequality holds as long as the numerator and denominator have the same sign, or $\tau > 2/3$. We therefore find that when $\tau > 2/3$, two open channels lead to higher noise than a single open channel. However, more than two channels can yield noise that is even smaller than noise produced by a single channel for $\tau > 2/3$, see Fig. S5. Note that in this figure, we divided the transmission equally between channels. However, in practice, at small conductance ($\tau < 1$) the third channel's contribution is typically significantly smaller than the second and first channels (as channels open approximately sequentially). Thus, it is a good approximation to associate the lower bound of the noise in the region $2/3 < \tau < 1$ with a single open channel model, rather than with an $N > 2$-channel model.

What about the region $0 < \tau \leq 1/3$? In this case we can satisfy (S44) when $\tau_1 \geq \tau(2-3\tau)/(1-3\tau)$, which is a contradiction to the basic requirement that $\tau_1 \leq \tau$. Thus, in this region as well, a single channel leads to higher noise than two channels.

In summary:

$$\mathcal{F} = \sum_i \tau_i^2(1-\tau_i) \quad \begin{cases} \text{is upper-bounded by a single channel for} & 0 \leq \tau \leq 2/3 \\ \text{is lower-bounded by a single channel for} & 2/3 < \tau < 1. \end{cases} \tag{S47}$$

## S5. FREQUENCY DEPENDENCE OF THE POWER SPECTRUM $\Phi_v(\omega)$

A popular mathematical model to explain the behavior of low-frequency noise, $S_f(f) \propto 1/f^\alpha$ is the Random Telegraph Signal (RTS)[10]. A single RTS has a Lorentzian frequency dependence in its power spectrum. In the time domain, the RTS noise shows switching of the current, voltage or resistance between two or more discrete values. In electronics, RTS is attributed e.g. to switching of configurations, reversible bond-breaking and forming, and other physical and chemical transformations[11]. A combination of several RTS can be shown to give rise to flicker noise with a power spectrum $S_f(f) \propto 1/f^\alpha$ in a certain frequency range (see e.g. Ref.[12]).

### A. Simulations of low frequency noise with random telegraph noise

Here we shall see how the Random Telegraphic Signal can be used to gain some insight about the time scales of scatterers involved, which may produce the experimentally observed $1/f$ frequency behavior of the flicker noise in atomic-scale junctions. Let us recall that the electric current depends on the transmission/reflection of the central ballistic region as well as the return amplitudes at the interface zones, as shown in Eqs. (S12) and (S13). The power spectral density was written as

$$S_f(\omega) \simeq 2G_0^2 V^2 \sum_i \mathcal{T}_i^2(1-\mathcal{T}_i) \sum_v \Phi_v(\omega), \tag{S48}$$

where $v = L, R$ labels the scattering region. The function

$$\Phi_v(\omega) = 2 \lim_{t_m \to \infty} \left| \int_0^{t_m} \boldsymbol{a}_v(\epsilon_F, t) e^{i\omega t} dt \right|^2 / t_m \tag{S49}$$

identifies contributions to the power spectral density due to fluctuations arising from the return amplitude $\boldsymbol{a_v}$. To eliminate confusion, in this Sec. we denote transmission probabilities by $\mathcal{T}$, instead of $\tau$, the latter is used to represent timescales related to scattering processes. In this section and in the numerical results presented, we include a prefactor of 2 in the definition of the power spectral density as a convention to include explicitly the contribution of the negative frequency part. We model each of those fluctuations in $\boldsymbol{a_{v,ii}}$ using a Random Telegraphic Signal and show how one may obtain a frequency dependence approximated by $S_f(f) \propto 1/f^\alpha$ for low frequencies, where $\alpha$ is close to one.

For a stationary process such as a Random Telegraph Signal, we may use the Wiener-Khinchin theorem to obtain an analytical expression for the power spectral density of Eq. (S49)

$$\Phi_{\boldsymbol{a_{v,ii}}}(\omega) = 2 \int_{-\infty}^{\infty} \langle \boldsymbol{a_{v,ii}}(t)\boldsymbol{a_{v,ii}}(t+t')\rangle e^{i2\pi f t'} dt',$$

where the angular brackets indicate long-time average.

The Random Telegraph noise consists of a signal switching between two values (also termed states or levels), which we denote by $\pm$. In this model, the fluctuating elastic scatterers modulate the return amplitudes $a$ between two values, $a_+$ and $a_-$. The Random Telegraph signal dwells in the + state (or level $a_+$) for an amount of time $t$ drawn from the distribution $\exp(-t/\tau_+)$, where $\tau_+$ denotes the mean dwell time for the upper level $a_+$. After it dwells for a time $t$ in the upper level, it switches to the lower level and dwells there for a time $t$ drawn from the distribution $\exp(-t/\tau_-)$ where similarly $\tau_-$ denotes the mean dwell time for the lower level $a_-$. We denote the difference between the amplitudes as $\Delta a = a_+ - a_-$. The expression in Eq. (S50) can be evaluated, and shown to be[13]

$$\Phi(f) = \frac{4(\Delta a)^2}{(\tau_+ + \tau_-)[(1/\tau_+ + 1/\tau_-)^2 + (2\pi f)^2]},$$

for a single RTS.

In Eq. (S48) we sum over noise contributions, $\Phi_v(f)$, from the two II regions. However, in each metal contact there are multiple independent scatterers that contribute to the $1/f$ noise, $\Phi(f) = \sum_v \sum_k \Phi_{v,k}(f)$. We assume that there are no fundamental differences between scatterers at the $L$ or $R$ sides, that is, for simplicity, they are characterized here by identical dwell times and amplitudes. As such, we relieve the summation over the regions $L$ and $R$ and directly count scatterers by $k$ without marking their position; we consider multiple ($M$) independent scatterers located in either sides, each of which has a Lorentzian spectral density given by Eq. (S51). The total power spectral density is given by $\Phi(f) = \sum_k \Phi_k(f)$

$$\Phi_k(f) = \frac{4(\Delta a_k)^2}{(\tau_{k_+} + \tau_{k_-})[(1/\tau_{k_+} + 1/\tau_{k_-})^2 + (2\pi f)^2]},$$

where the index $k$ represents the contribution due to a scatterer $k$; we assume that parameters here are independent of the channel $i$. In Figs. S6-S8 we plot $S_f(f) = 2G_0^2V^2 \sum_i \mathcal{T}_i^2(1-\mathcal{T}_i)\sum_k \Phi_k(\omega)$.

In order to make the notation more concrete, consider one scatterer on the right bank, enumerated as '1'. The return amplitudes fluctuate between two values, $a_{1_+}$ and $a_{1_-}$. The value $a_{1_+}$ is maintained for a mean dwell time given by $\tau_{1_+}$ before switching to the lower value. Similarly, the lower level persists for a mean dwell time given by $\tau_{1_-}$ before switching back. We denote the difference between the return amplitudes by $\Delta a_1 = a_{1_+} - a_{1_-}$.

In simulations presented here, we take $\Delta a_k = \Delta a$ to be the same for all scatterers. We also assume equal dwell times for the two states of each scatterer, $\tau_{k_+} = \tau_{k_-} = \tau_k$. If there are $M$ scatterers present, the index $k = \{1,2,3,...,M\}$. Note here that we work under the assumption that each scatterer affects the return amplitudes in every channel in the same way.

In Figs. S6 and S7, we consider a single transmission channel and two scatterers with well-separated time scales $\tau_2 = 10\tau_1$. These figures exemplify the impact of the mean dwell times on the power spectrum. In Fig. S8, we consider two transmission channels and a total of three scatterers, all with well-separated timescales $\tau_3 = 10\tau_2 = 100\tau_1$. In this example we further add white noise to the current signal, demonstrating that it does not affect the $1/f$ region.

In more details, in Figs. S6 and S7, we study a model with a single conducting channel with transmission $\mathcal{T} = 0.7$ and a voltage bias of $V = 1$ mV. Using $\langle I \rangle = \frac{2e^2}{h}\mathcal{T}V$, this results in a steady-state current of about $\langle I \rangle \simeq 54$ nA. We take into account two independent RTSs (say one positioned at the left bank and one at the right), which give rise to current fluctuations via fluctuations in the return amplitudes $\boldsymbol{a}_k(t)$, $k = \{1,2\}$. However, under our assumptions, the results will be unchanged even if both scatterers were on the same region. Changes in the return amplitudes modify the transmission, as can be seen from Eq. (S12). The difference between the two levels of the return amplitude has been set to $\Delta a = 10^{-3}$, which is small compared to the steady state transmission $\mathcal{T} = 0.7$. Fluctuations in the return amplitude lead to fluctuations in the current, and for a combination of two RTSs, the current fluctuates between three values. These correspond to both scatterers being in the up state, one scatterer in the up and the other in the down state, and finally both scatterers in the down state. Note that (up, down) and (down, up) for the first and second scatterers, respectively give the same current since $\Delta a_1 = \Delta a_2 = 10^{-3}$. The fluctuations in the current are shown in Figs. S6b and S7b with respect to the mean steady-state current.

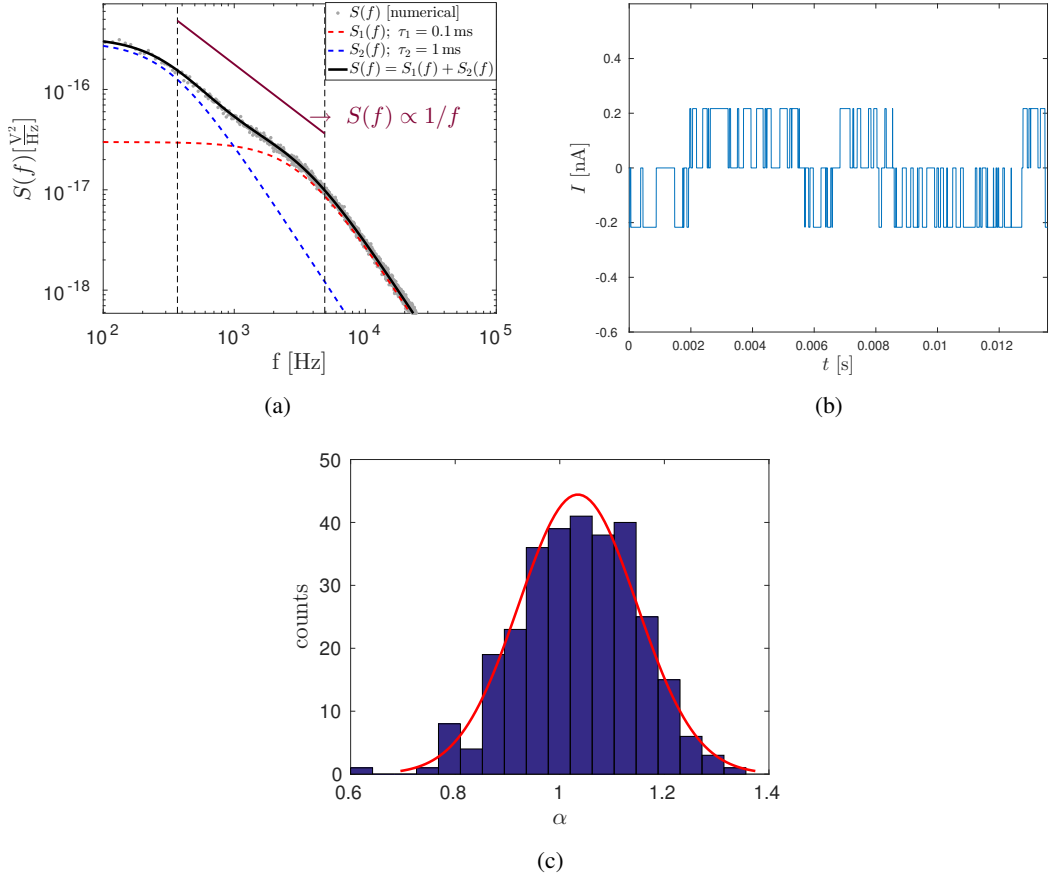

(a)

(b)

(c)

Figure S6: (a) Addition of two independent RTS. Mean dwell times of the two RTSs are $\tau_{1+} = \tau_{1-} \equiv \tau_1 = 0.1\,\mathrm{ms}$ and $\tau_{2+} = \tau_{2-} \equiv \tau_2 = 1\,\mathrm{ms}$. The analytical spectra of each RTS is shown with dashed red and blue curves respectively. The solid black line shows their (analytically calculated) sum, and agrees well with the numerically calculated spectrum shown in gray. We use $\Delta a = 10^{-3}$ for both RTSs. Purple line is $S = A/f^{\alpha}$, with slope $\alpha = 1$ and some $A$. Black dashed vertical lines in (a) show the region of the curve that was fitted to $S = A/f^{\alpha}$. (b) Current trajectory $I(t)$ with respect to the steady state current $\langle I \rangle \simeq 54\,nA$. (c) Histogram of slopes $\alpha$ can be seen to be centered around 1. Mean slope $\bar{\alpha} = 1.02$ and the standard deviation $\Delta\alpha = 0.096$. Ensemble average taken over 300 trajectories.

In Fig. S8, we consider three scatterers but still a single transmission channel with transmission probability $\mathcal{T} = 0.7$ and the same bias condition. The steady-state current is therefore the same, $\langle I \rangle \simeq 54\,nA$. The three independent scatterers are labeled by $k = \{1, 2, 3\}$. For concreteness, we may say that we have two scatterers on the left and one on the right. The level separation for the return amplitudes are the same for each scatterer $\Delta a_1 = \Delta a_2 = \Delta a_3 = 10^{-3}$.

### B. Appearance of $1/f$ behavior

The time scales of the RTS is defined by the mean dwell times $\tau_{k-} = \tau_{k+} = \tau_k$ for the lower and upper levels, which we here take to be equal. When we consider two such RTSs having well-separated time scales, we start to see the appearance of the $1/f$ behavior in the power spectral density. In Fig. S6, the time scales for the two RTSs are taken to be $\tau_1 = \tau_{1+} = \tau_{1-} = 0.1$ ms and $\tau_2 = \tau_{2+} = \tau_{2-} = 1$ ms. The $1/f$ frequency behavior is seen in a frequency window shown by the black vertical lines in Fig. S6a. We find an excellent agreement between numerical results (shown in gray), with that of the analytical spectrum given by Eq. (S52) (shown in black). The dashed red and blue curves correspond to the analytically calculated spectral densities of the individual RTSs. The black curve is their sum. Numerical results are obtained by generating 300 different RTS spectra, each for a total time which is much longer than the longer mean dwell times, $T >> \tau_2$ (typically 100 times longer). In the numerical simulations of Figs. (S6) and (S7), the total time for the current trajectory is taken to be $\simeq 65$ ms, which is many times longer than the mean dwell times.

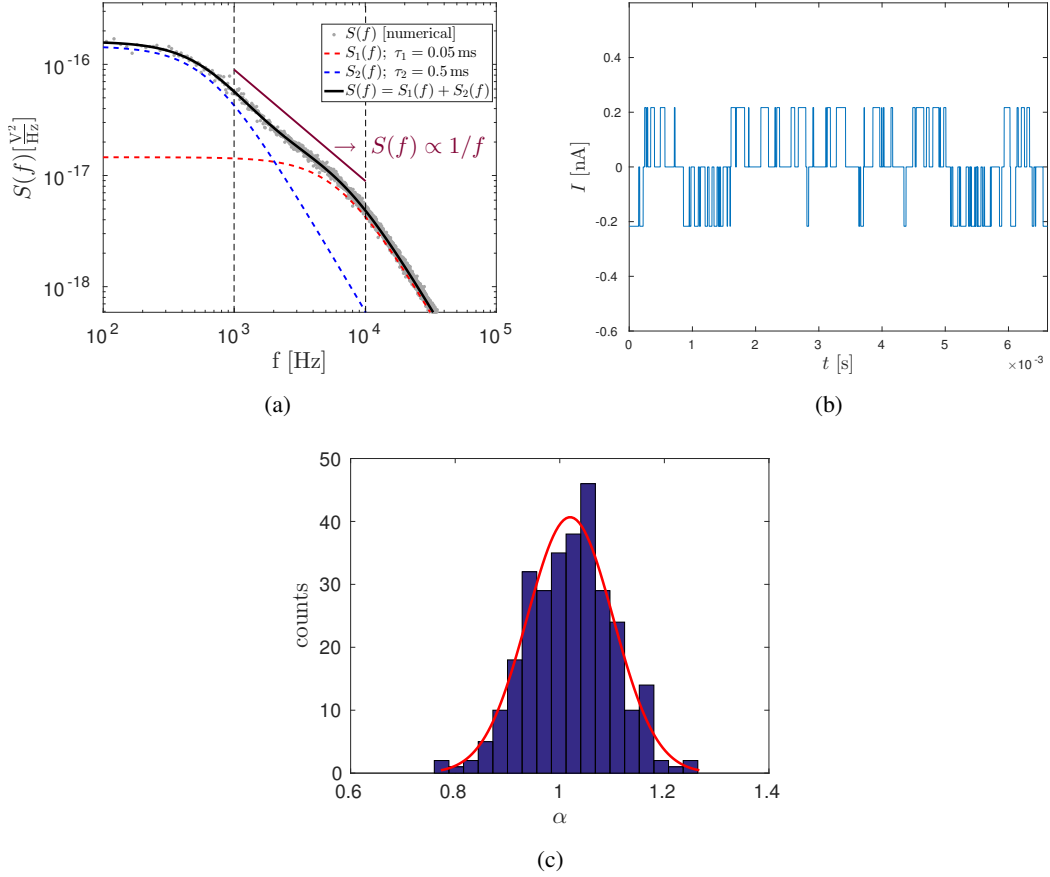

(a)

(b)

(c)

Figure S7: (a) Addition of two independent RTS with short dwell times. Notice the effect of decreasing the mean dwell times is to move the Lorentzian spectra to the right, towards higher frequencies. Mean dwell times of the two RTSs are $\tau_1 = \tau_{1+} = \tau_{1-} = 0.05\text{ms}$ and $\tau_2 = \tau_{2+} = \tau_{2-} = 0.5\text{ms}$. The analytical spectra of each RTS is shown with dashed red and blue curves respectively. The solid black line shows their (analytically calculated) sum, and agrees extremely well with the numerically calculated spectrum shown in gray. We use $\Delta a = 10^{-3}$ for both RTSs. Purple line is $S = A/f^\alpha$, with slope $\alpha = 1$ and some $A$. Black dashed vertical lines in (a) show the region of the curve that was fitted to $S = A/f^\alpha$. (b) Current trajectory $I(t)$ is shown with respect to the steady state current $\langle I \rangle \simeq 54 \ nA$. (c) Histogram of slopes $\alpha$ can be seen to be centered around 1. Mean slope $\bar{\alpha} = 1.03$ and the standard deviation $\Delta\alpha = 0.09$. Ensemble average taken over 300 trajectories.

Numerically, the power spectral density is computed not by using the Wiener-Khinchin theorem but by directly calculating the Fourier transforms of the current trajectory as given by $S(\omega) = 2 \lim_{t_m \to \infty} \frac{1}{t_m} \left| \int_0^\infty \delta I(t) e^{i\omega t} dt \right|^2$, where we have now included a prefactor of 2 for Eq. (S15) as per our convention. The slope $\alpha$ in the resulting power spectral density $S \propto 1/f^\alpha$ is calculated for each of these trajectories and its histogram is shown in Figs. S6b, S7b and S8b. In Fig. S7, the mean dwell times for both the RTSs are cut by half $\tau_{1+} = \tau_{1-} = 0.05$ ms and $\tau_{2+} = \tau_{2-} = 0.5$ ms, compared to Fig. S6. The reduction of the dwell time pushes the Lorentzian spectra towards the right, higher frequencies region. The time scales chosen in Fig. S7 show very good $1/f$ behavior between $10^3$ and $10^4$ Hz, which corresponds to the experimental window of observation. One may therefore gets a picture for the type of timescales at which the scatterers in the experiment are operating.

In Fig. S8, we introduce an additional scatterer for a total of three. We find that the $1/f$ behavior has extended in its frequency range by introducing this additional scatterer. The three scatterers are again well-separated in terms of their mean dwell times, $\tau_1 = \tau_{1+} = \tau_{1-} = 0.01$ ms ,$\tau_2 = \tau_{2+} = \tau_{2-} = 0.1$ ms and $\tau_3 = \tau_{3+} = \tau_{3-} = 1$ ms. We also introduce here an additional white noise component for each of these scatterers by using a Gaussian distribution whose standard deviation is set to $0.25 \Delta a$. We observe in the power spectral density, Fig. S8a that there is a discrepancy between the numerical and analytical spectra in the higher frequency region. The white noise contribution starts to appear in the high frequency region (about $10^5$ Hz) where the $1/f$ contribution becomes quite small. The added white noise does not affect the lower frequency region since it is dominated

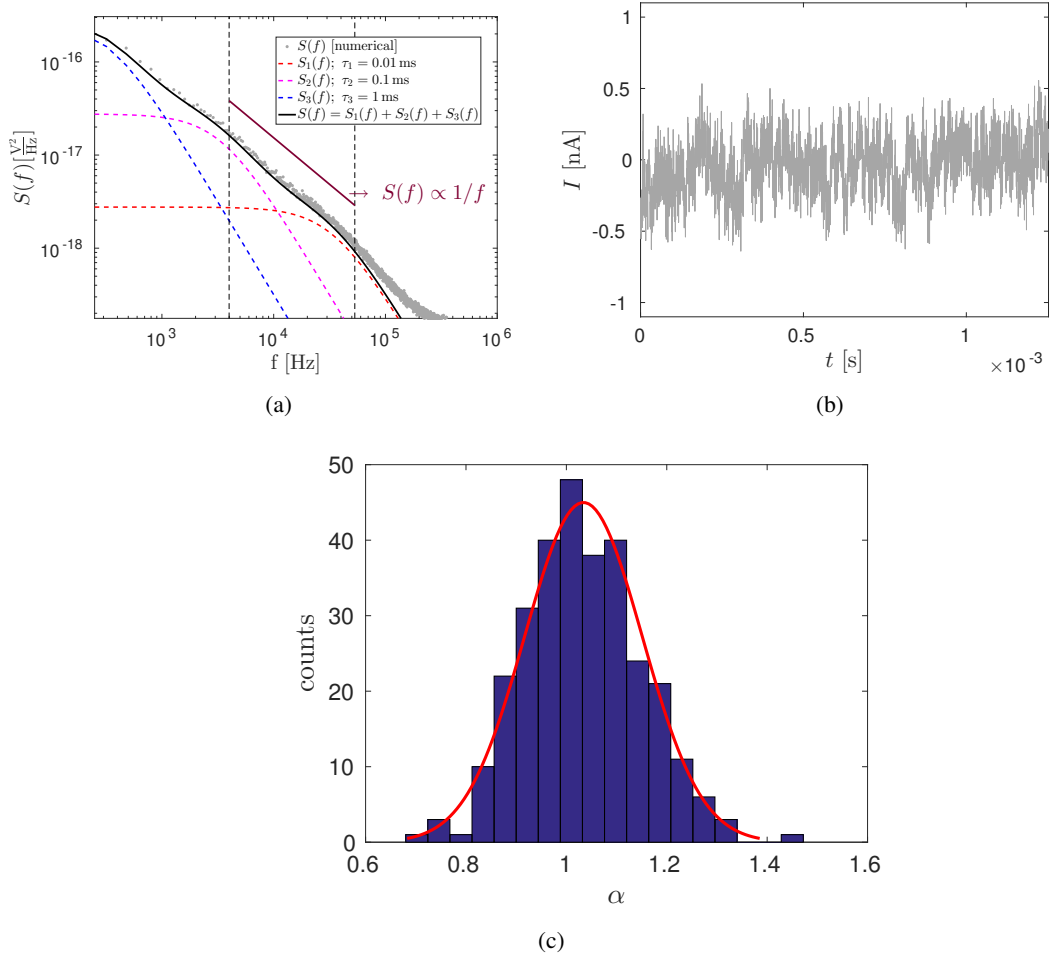

(a)

(b)

(c)

Figure S8: (a) Addition of 3 independent RTS with an added white noise. Including more scatterers with well-separated timescales (defined by their mean dwell times) serves to elongate the region in which we observe the $1/f$ behavior. Mean dwell times of the two RTSs are taken to be $\tau_1 = \tau_{1+} = \tau_{1-} = 0.01\,\mathrm{ms}$, $\tau_2 = \tau_{2+} = \tau_{2-} = 0.1\,\mathrm{ms}$ and $\tau_{3+} = \tau_{3-} \equiv \tau_3 = 1\,\mathrm{ms}$. The analytical spectra of each RTS is shown with dashed red, magenta and blue curves respectively. The solid black line shows their (analytically calculated) sum, and agrees well with the numerically calculated spectrum shown in gray. The added white noise only affects the high frequency part of the spectral density; here one can see the departure between the analytically calculated spectrum and the numerical one. We use $\Delta a = 10^{-3}$ for all the three independent RTSs. Purple line is $S = A/f^\alpha$, with slope $\alpha = 1$ and some $A$. Black dashed vertical lines in (a) show the region of the curve that was fitted to $S = A/f^\alpha$. (b) Current trajectory $I(t)$ with respect to the steady state current $\langle I \rangle \simeq 54\ nA$. (c) Histogram of slopes $\alpha$ can be seen to be centered around 1. Mean slope $\bar{\alpha} = 1.02$ and the standard deviation $\Delta\alpha = 0.11$. Ensemble average taken over 300 trajectories.

by the flicker noise.